\documentclass[12pt,letterpaper,oneside]{lsuthesis}
\usepackage{bm}
\usepackage{bbold}
\usepackage[margin=1in]{geometry}
\usepackage{braket}
\usepackage{graphicx}
\usepackage{amsmath}
\usepackage[boxed, vlined]{algorithm2e}
\usepackage{doi}
\usepackage{setspace}
\usepackage[numbers,sort&compress]{natbib}

\hypersetup{%
  pdftitle={My Thesis},
  pdfauthor={Nathaniel Robert Miller},
  pdfkeywords={Physics,Astronomy},
  pdfdisplaydoctitle=true,
  bookmarksnumbered,
  pdfstartview={FitH},
  colorlinks,
  citecolor=black,
  filecolor=black,
  linkcolor=black,
  urlcolor=black,
  breaklinks=true,
  hypertexnames=false
}

\newcommand{\pare}[1]{\left( #1 \right)}
\newcommand{\ave}[1]{\langle #1 \rangle}
\newcommand{\sinc}{\text{sinc}}
\newcommand{\abs}[1]{\left\vert #1 \right\vert}
\newcommand{\cor}[1]{\left[ #1 \right]}
\begin{document}
\pagenumbering{roman}
%
\pagestyle{empty}
\begin{center}%

  {\fontsize{16}{19}\selectfont\bfseries NEW ASPECTS OF OPTICAL COHERENCE AND THEIR POTENTIAL FOR QUANTUM TECHNOLOGIES \par}%

  \vfill

  {\normalsize A Dissertation\par}
  \vskip 1em
  {\normalsize Submitted to the Graduate Faculty of the \par
    Louisiana State University and \par
    Agricultural and Mechanical College \par
    in partial fulfillment of the \par
    requirements for the degree of \par
    Doctor of Philosophy\par}
  \vskip 1em
  {\normalsize in\par}
  \vskip 1em
  {\normalsize The Department of Physics and Astronomy
    \par}

  \vfill

  {\normalsize by \par}
    {\normalsize Nathaniel Robert Miller\par}
    {\normalsize B.S., University of Dayton, 2018 \par}%
    {\normalsize May 2022 \par}%

\end{center}\par
%
\clearpage 
\pagestyle{plain}
%
%
\clearpage
\vspace*{\fill}
\protect
\protect
\begin{center}
I dedicate this work to the memory of Jonathan Dowling
\end{center}
\protect
\protect
\vspace{\fill}
%
%
\doublespacing
\clearpage
%
\chapter*{\acknowledgmentname}
%
\addcontentsline{toc}{chapter}{\acknowledgmentname}
\protect
\protect
First and foremost I would like to thank my original advisor Dr. Jonathan P Dowling.  Through my early days at LSU he was always available for questions, and checking to make sure I was okay.  He gave me the chance to travel and access collaborations that were essential for me becoming the physicist I am today.

I would like to thank my advisor Dr. Omar Maga\~na-Loaiza for offering to become my advisor after Jon passed. His mentorship and experimental knowledge allowed me to grow as a theorist, and gave me a unique perspective that I will carry with me throughout my career.

I would like to Dr. Chenglong You for his help in the past two years.  Thanks to your availablity and knowledge of both theory and experimental physics I was able to improve upon my own knowledge and better improve my knowledge of both.

I am grateful to my collaborators, Dr. Peter P Rohde; Dr. Roberto de J. Le\`on Montiel for their advice and support throughout my PhD.
I am grateful for my committee, Dr Hwang Lee; Dr William Shelton and Dr. David Koppelman for agreeing to participate.

I would like to thank my colleges, Pratik Barge, Arshog Danageozian and Dr. Narayan Bhusal for their support and friendship throughout my PhD.

I thank the faculty at the Univeristy of Dayton, Dr. William N. Plick, Dr. Leno P. Pedrotti, Dr. Jay Matthews and Dr. Said Elhamri for their inspiration and support throughout my undergraduate degree and for leading me down this path.  

I would like to thank my friends, Alex Igl, Ali Crisp, Courtney Crawford.  Without any of them I would never have made this far.  I am the person I am today thanks to their friendship, support and willingness to be yourselves.

I would also like to acknowledge financial support from the LSU Physics and Astonomy department without which this work would have never been possible.

Lastly I would like to thank my parents, Dr. Michael E Miller and Karen Miller.  Without their support throughout my childhood and beyond.  I wouldn't have made it this far without you.

\protect
\protect

\singlespacing
\tableofcontents

%

%
%

\clearpage
\singlespacing
\chapter*{\abstractname}
%
\addcontentsline{toc}{chapter}{\abstractname}
\vspace{-1em}
\protect
\protect
Currently, optical technology impacts most of our lives, from light used in scientific measurement to the fiber optic cables that makeup the backbone of the internet. However, as our current optical infrastructure grows, we discover that these technologies are not limitless.  Astronomers find themselves unable resolve stars that are too close to one another.  Meanwhile, the internet is always under threat as our computer technology improves and more complex ways to break encryption emerge, threatening our personal information and infrastructure.  However, our current optical technology functions on classical principles, and can be easily improved by incorporating our knowledge of quantum optics. In order to implement quantum technologies, our understanding of quantum coherence must improve.  Through this knowledge we can maintain quantum states, and therefore their information, longer.  In this dissertation, I will demonstrate that with sufficient knowledge of coherent properties, a simple algebra can be derived which can provide rules for graph reductions on a quantum network graph.  Using this knowledge, I then provide a rudimentary algorithm which can find the optimal subgraph for communication on a quantum network.  Next, I demonstrate that by measuring the photon statistics and second-order quantum coherence of a field, one can create a neural network capable of distinguishing the light sources on a pixel.  Which is then applied to develop an imaging scheme capable of surpassing the Abbe-Rayleigh Criterion. Lastly, I present a multiphoton quantum version of the van Cittert-Zernike theorem.  This provides formalism capable of determining the propagation of quantum coherence throughout a system.  I then demonstrate the usefulness of the theorem by demonstrating sub-Poissonian statistics created by a linear system with an incident thermal beam, obtainable only by post-selection. Altogether, this provides incite into new applications of coherence to quantum technologies and the formalism to extending our knowledge even further.
\doublespacing
\protect
\protect

%
\mainmatter
\pagenumbering{arabic}
%
\singlespacing
\chapter*{Chapter 1. \\ Introduction to Quantum Technologies}
\addcontentsline{toc}{chapter}{Chapter 1. Introduction to Quantum Technologies}
\setcounter{chapter}{1}
\setcounter{section}{0}
\setcounter{figure}{0}
\setcounter{table}{0}
\label{Intro}
\doublespacing
Optical technologies form the backbone of modern day society. This can be seen in microscopy and astronomy where making increasingly precise measurements is essential for advancing our understanding \cite{ferraro2011coherent,paur2018tempering,parniak2018beating,bib:Dowling08}. It is present every time we use our phones or our computers as fibre optic cables are essential for the creation of the internet. However, as our measurements become more precise, and the internet becomes a larger part of our day-to-day life, we begin to push against the boundaries set by our current technology, since they are limited by classical phenomena \cite{rayleigh1879xxxi,Saleh18Optica,rohde2021quantum,bib:xx,bib:kimble2008quantum}. This suggests that it's time to start on a new paradigm, one with an expected impact similar to the invention of the silicon chip or the modern day internet. This new paradigm comes in the forms of quantum technologies, technologies not limited by classical physics and instead limited by fundamental limits imposed by the universe \cite{hell20152015,gerry2005introductory,bib:Dowling08,bib:Deutsch85}. 

Therefore, the continued investigations into the nature of quantum coherence is essential to the future of technology. Increased knowledge of coherence can allow for either the exploitation or avoidance of coherence effects on the desired system. Allowing for the creation of better memories, gates and sensors. In this dissertation, I present a variety of models and applications, ultimately improving our knowledge of coherence. In Chapter \ref{Quantum Networks}, I present a simple model of an entanglement network were I aim to find the optimal set of channels to transmit data while keeping the qubits coherent. I demonstrate in the case of dephasing, one can derive a set of algebraic rules that ultimately provide the optimal subgraph that information can be transmitted through \cite{leone2021qunet}. In Chapter \ref{QI}, I demonstrate that our current knowledge of quantum statistics and quantum coherence gives the ability to describe the interaction of multiple states. Then, exploiting that information, I demonstrate that a machine learning algorithm can use this information to distinguish light sources on a single pixel, enabling the creation of a camera that can surpass the Abbe-Rayleigh criterion \cite{bhusal2021smart}. Lastly, in Chapter \ref{MQvCZT}, I demonstrate a multiphoton quantum version of the van Cittert-Zernike theorem. This enables a model capable of measuring the propagation of coherence properties throughout the system. Furthermore, I apply this model to a simple linear system acted on by an unpolarized thermal beam, demonstrating that through certain post-selected measurements it is possible to take sub-shot noise measurements, demonstrating the power of such a tool \cite{miller2022multiphoton}. 

In the remainder of this chapter I briefly review quantum optics and quantum technologies. In Chapter \ref{Coherence}, I provide additional background on both quantum and classical coherence, allowing the reader to understand the basis of this dissertation. 

\section{Quantum Technolgies}
\label{Intro:QT}

Quantum sensing, quantum communication and quantum information processing together all form the basis of quantum technologies. These systems can all be implemented using light sources that are either in the low intensity or single-photon regimes, though many other platforms exist \cite{devoret2004superconducting,doi:10.1063/1.5089550,doi:10.1126/science.1231298}. As such these systems are extremely effected by environmental effects, which can result in loss or another phenomena known as decoherence \cite{bib:Shor95,bib:NielsenChuang00,Perrier_2020,steane1998space,PhysRevA.64.032101}. Decoherence is currently one of the largest hurdles in implementing real life quantum devices, particularly in the fields of quantum communication and information processing \cite{pellizzari1995decoherence,steane1998quantum,yuan2010entangled,BRANDT1999257}. This stems from the fact that both applications require quantum states to stay coherent for as long as possible \cite{Quiroz-Juarez:19,Tschernig:18}. Furthermore, both of these applications have both required and unavoidable interactions with the quantum state with outside sources \cite{steane1998quantum,bib:PhysRevLett.95.250503,PhysRevB.72.014417}. In the case of quantum information processing, one major source of decoherence is the action of gates. While on paper gates seem not to present an issue, real world applications require coupling the state to a much larger system, allowing the state to leave its ideal subsystem while it is manipulated, leaving the possibility that it will not return \cite{fedorov2012implementation,waseem2015realization,niu2019universal}. Consequently, the state can also return to the subsystem and be unable to maintain the correct value \cite{fedorov2012implementation,niu2019universal,stricker_experimental_2020}. While communications also has to deal with problems of gates and memory in order to perform essential functions, they also are plagued by environmental noise as the data is transmitted \cite{kurtsiefer2002step,gabay2006quantum,wang2018atmospheric,bib:NJP_11_075001}. While fortunately, photons aren't subject to nearly as much decoherence as most platforms, their inability to stay in place increases the likelihood of loss \cite{o2007optical,knill2001scheme,lvovsky2009optical}. Furthermore, they lack an easy form of memory, causing the need for long, noisy delay lines to act as memory, or the use of another quantum platform to store information \cite{tanabe2007trapping,maitre1997quantum,leung2006quantum,danageozian2021noisy,bib:RalphHayes05,lvovsky2009optical,bib:chen2013coherent,bib:jin2015telecom}. Each of these introduces additional chances of decoherence and loss that must be corrected or avoided in order to create a fully functioning quantum system \cite{o2007optical,knill2001scheme,tanabe2007trapping,maitre1997quantum,leung2006quantum,danageozian2021noisy,bib:RalphHayes05,lvovsky2009optical,bib:chen2013coherent,bib:jin2015telecom}.

Quantum sensing on the other hand looks to exploit optical coherence in order to create better more precise measurements \cite{lemos2014quantum,Miller2021versatilesuper,bib:Dowling08,tsang2009quantum,dorner2009optimal}. While the sensitivity of quantum states ultimately proves to be troubles when trying to building systems isolated from the environment, this "supersensitivity" is ideal for sensing where stronger interactions are often desirable. Two of the most interesting applications in sensing are the Hong-Ou-Mandel and Induced Coherence experiments \cite{hong1987measurement,zou1991induced}. The Hong-Ou-Mandel experiment functions as a quantum eraser, two paths of light proceed through the system but cannot be distinguished from one another at measurement \cite{hong1987measurement}. The complete lack of which-path information leads to know interference between the two beams, suggesting that they are coherent. However, as path information is introduced in the form of a polarizer placed into one of the paths, interference begins to occur. This allows information about the polarization properties of an object placed into one of the arms to be created by measuring the destruction of coherence in the form of available interference. The Induced Coherence experiment performed originally by Zou, Wang and Mandel, functions similarly at first glance, relying primarily on the destruction of coherence by creating path information in order to make a measurement \cite{zou1991induced,Kolobov_2017,lemos2014quantum,Miller2021versatilesuper}. However, as discovered by the experiment, coherence can be destroyed through paths not involved in the final measurement. Allowing for measurements to be taken with a light beam that doesn't even interact with the object in question. While these experiments form the early basis of quantum sensing, they have since been expanded to a wide variety of applications, such as imaging, remote sensing, gravitational wave astronomy, or many other applications\cite{lemos2014quantum, lahiri2015theory,tsang2016quantum,Steinberg17PRL,zhou2019quantum,Treps20Optica,Saleh18Optica,Liang21Optica,Miller2021versatilesuper,kumar2021single, huver2008entangled,okane2021quantum,zhuang2017entanglement,dorner2009optimal, schnabel2010quantum}.


\section{Quantization of the Electromagnetic field}
\label{Intro:Quant}

In order to exceed limits in classical optics quantum model of light must be created. I begin by quantizing the electromagnetic field. The classical description of electromagnetic fields with no sources of radiation is governed by Maxwell's Equations \cite{zangwill2013modern}:
\begin{equation}\label{Chap1:Eqn:Maxwell's}
\begin{aligned}
\nabla\times\mathbf{E}=-\frac{\partial \mathbf{B}}{\partial t},\\
\nabla\times\mathbf{B}=\mu_0\epsilon_0\frac{\partial\mathbf{E}}{\partial t},\\
\nabla\cdot\mathbf{E}=0,\\
\nabla\cdot\mathbf{B}=0.
\end{aligned}
\end{equation}
\noindent
Where $\mathbf{E}$ is the electric field vector, $\mathbf{B}$ is the magnetic field vector, $\epsilon_0$ is the permittivity of free space and $\mu_0$ is the permeability of the vacuum. We can then consider the field inside a perfectly conducting cavity with length L. Let the electric field be x-polarized such that $\mathbf{E}=\mathbf{e}_x E_x$, where $\mathbf{e}_x$ is the polarization unit vector. Allowing for only a single mode field, the solution to Maxwell's equations are \cite{gerry2005introductory}
\begin{equation}\label{Chap1:Eqn:fields}
\begin{aligned}
E_x(z,t)=\left(\frac{2\omega^2}{V\epsilon_0}\right)^{\frac{1}{2}}q(t)\sin(kz),\\
B_y(z,t)=\left(\frac{\mu_0\epsilon_0}{k}\right)\left(\frac{2\omega^2}{V\epsilon_0}\right)^{\frac{1}{2}}\dot{q}(t)\cos(kz).
\end{aligned}
\end{equation}
\noindent
Where $k$ is the wavenumber of the field, $\omega$ is the frequency of the field with the allowed values defined as $\omega_m=c(m\pi / L)$, and V is the colume of the cavity. Let $q(t)$ be a time dependant factor with units of length and $\dot{q}(t)$ its first derivative in time. The total Hamiltonian of our electromagnetic field is therefore

\begin{align}
H=\frac{1}{2}\left(\dot{q}^2+\omega^2 q^2\right).
\end{align}

\noindent
Here, $q$ is the canonical position and $\dot{q}$ is the canonical momentum. This gives the same Hamiltonian as the simple harmonic oscillator. Therefore, the commutation relation between the position and momentum operator is defined as $[\hat{q},\hat{p}]=i\hbar \hat{I}$ where I set $\dot{q}=p$ to better represent it as momentum. Substituting the position and momentum operators into their function equivalents in Eqn. (\ref{Chap1:Eqn:fields}) returns the electric and magnetic field operators. The ladder operators can then be defined as

\begin{equation} \label{chap1:eqn:LadOps}
\begin{aligned}
\hat{a}=\left(2\hbar\omega\right)^{-1/2}\left(\omega\hat{q}+i\hat{p}\right),\\
\hat{a}^{\dag}=\left(2\hbar\omega\right)^{-1/2}\left(\omega\hat{q}-i\hat{p}\right).
\end{aligned}
\end{equation}

\noindent 
Note that the ladder operators follow the commutation relation $\left[\hat{a},\hat{a}^{\dag}\right]=\hat{I}$. The field operators can then be written in terms of the ladder operators as
\begin{equation}\label{Chap1:Eqn:Fields}
\begin{aligned}
\hat{E}_x\left(z,t\right)=\mathcal{E}_0\left(\hat{a}+\hat{a}^\dagger\right)\sin\left(kz\right), 
\end{aligned}
\end{equation}
\begin{equation}
\begin{aligned}
\hat{B}_y\left(z,t\right)=-i\mathcal{B}_0\left(\hat{a}-\hat{a}^\dagger\right)\cos\left(kz\right).
\end{aligned} 
\end{equation}

\noindent 
Where $\mathcal{E}_0=\left(\hbar\omega/\epsilon_0 V\right)^{1/2}$ and $\mathcal{B}_0=\left(\mu_0/k\right)\left(\epsilon_0\hbar\omega^3/V\right)^{1/2}$ which represents the amount of electric and magnetic field per photon. It is also useful to define the positive component of the electric field as
\begin{equation}\label{Chap1:Def:Efield}
\begin{aligned}
\hat{E}^{(+)}=\frac{i}{\sqrt{2}}\mathcal{E}_0\hat{a}.
\end{aligned} 
\end{equation}
Consequently, the negative field is the complex conjugate of the positive field. Using Eqn, \ref{chap1:eqn:LadOps} the Hamiltonian can be re-written as

\begin{equation}\label{chap1:eqn:quanHam}
\begin{aligned}
\hat{H}=\hbar\omega\left(\hat{a}^{\dag}\hat{a}+\frac{1}{2}\right).
\end{aligned}
\end{equation}

It's important to note that $\hat{a}^{\dag}\hat{a}$ is the number operator $\hat{n}$ representing the number of photon is the field. Therefore, the zero point energy our system is $E_0=\hbar\omega/2$. The eigenstates of the hamiltonian form the Fock basis where $\hat{H}\ket{n}=E_n\ket{n}=\hbar\omega\left(n+\frac{1}{2}\right)\ket{n}$ where n is the photon number. Therefore, each state vector represents a number state, which is an uncommon quantum state. However, taking advantage of the number states forms a complete basis will make it essential for decomposing other states of light into it in order to determine their statistics.

\section{Thermal States}
\label{Intro:Thermal}

 Thermal light was originally described in any detail by Planck's radiation law attempting to describe blackbody radiation. Traditionally, blackbody radiation is emitted by a cavity made up of a material that is a perfect emitter and absorber \cite{gerry2005introductory}. This cavity contains thermal radiation at thermal equilibrium with its walls, and is coupled to a heat bath. This keeps it from being a truly free field and significantly alters its properties. In reality, thermal light describes all forms of blackbody radiation, such as sunlight, starlight or light from an incandescent bulb. Due to its (often unwanted) prevalence it is an essential state to understand for any application of quantum systems.

It is important to note that due to the presence of a heat bath, thermal states cannot be described by a state vector, instead we have to move our description to a density matrix. This allows a full description of the state through its description as a mircrocanonical ensemble. That the probability of a single mode field being excited to the nth level is given as \cite{pathria2016statistical}

\begin{equation} \label{Chap1:Eqn:ThermProb}
\begin{aligned}
P_n=\frac{\text{exp}\left(-E_n/k_b T\right)}{\sum_n\text{exp}\left(-E_b/k_bT\right)}.
\end{aligned}
\end{equation}
Where $T$ is the temperature of the blackbody and $k_b$ is the Boltzmann constant. The density matrix is constructed by taking the mircocanonical description and the Hamiltonian given by Eqn. (\ref{chap1:eqn:quanHam}) giving
\begin{equation}\label{Chap1:Eqn:ThermDensOne}
\begin{aligned}
\hat{\rho}_{\text{Th}}=\frac{\text{exp}\left(-\hat{H}/k_B T\right)}{\text{Tr}\left[\text{exp}\left(-\hat{H}/k_B T\right)\right]}.
\end{aligned}
\end{equation}
Where $\text{Tr}\left[\dots\right]$ is the trace. Defining $Z=\text{Tr}\left[\exp\left(\hat{H}/k_b T\right)\right]$, where Z is the partition function. The density matrix can be written in the fock basis by first noting
\begin{equation}\label{Chap1:Eqn:ThermProb2}
\begin{aligned}
 P_n=\bra{n}\hat{\rho}_{\text{Th}}\ket{n}=\frac{1}{Z}\text{exp}\left(-E_n/k_BT\right).
\end{aligned}
\end{equation}
Which is the same as the probability given in Eqn. (\ref{Chap1:Eqn:ThermProb}). The average photon $\bar{n}$ to be
\begin{equation}\label{Chap1:Eqn:aveNTherm}
\begin{aligned}
 \bar{n}=\frac{1}{\exp\left(\hbar\omega/k_B T\right)-1}\approx\exp\left(-\hbar\omega/k_B T\right). 
\end{aligned}
\end{equation}
Where $k_B T>>\hbar\omega$. Applying the approximation to Eqn. (\ref{Chap1:Eqn:aveNTherm}) and Eqn. (\ref{Chap1:Eqn:ThermProb2}) allows the density matrix to be re-writen as
\begin{equation}\label{Chap1:Eqn:ThermFinal}
\begin{aligned}
 \hat{\rho}_{\text{Th}}=\sum_n \frac{\left(\bar{n}\right)^n}{\left(1+\bar{n}\right)^{n+1}}\ket{n}\bra{n}.
\end{aligned} 
\end{equation}
The variance of a thermal state can then be noted to be $\langle\left(\Delta n\right)^2\rangle=\bar{n}^2+\bar{n}$, which will be important for state identification and coherence. These will provide a fundamental state for chapters \ref{QI} and \ref{MQvCZT}, where in Chapter \ref{QI} they will serve as one of the two states we hope to distinguish. In Chapter \ref{MQvCZT}, I will use a two-mode thermal state to describe the unpolarized, incoherent beam at the start of the system. 
\section{Displaced States and the Displacement Operator}
\label{Intro:Coh}
The next state of interest is the eigenstates of the annihilation operator originally defined by Eqn. (\ref{chap1:eqn:LadOps}), which can be found by solving the equation \cite{agarwal2012quantum}
\begin{equation}\label{Chap1:Eqn:CohStateEq1}
\begin{aligned}
 \hat{a}\ket{\alpha}=\alpha\ket{\alpha}.
\end{aligned}
\end{equation}
The eigenstate $\ket{\alpha}$ can be decomposed into the Fock basis yielding
\begin{equation}\label{Chap1:Eqn:NumDecomp}
\begin{aligned}
 \ket{\alpha}=\sum_n C_n\ket{n},
\end{aligned} 
\end{equation}
since the Fock basis is complete. Plugging Eqn. (\ref{Chap1:Eqn:NumDecomp}) into Eqn. (\ref{Chap1:Eqn:CohStateEq1}) gives
\begin{equation}
\begin{aligned}
\sum_n C_n\sqrt{n}\ket{n-1}=\alpha\sum_n C_n\ket{n}. 
\end{aligned}
\end{equation}
Equating the coefficients on both sides and solving yields
\begin{equation}
\begin{aligned}
\ket{\alpha}=C_0\sum_n \frac{\alpha^n}{\sqrt{n!}}\ket{n}.
\end{aligned} 
\end{equation}
Where $C_0$ is the weight of the vacuum state. Noting $\langle\alpha|\alpha\rangle=1$ and solving for $C_0$ giving us our final normalized state of
\begin{equation}\label{Chap1:Eqn:CoherentState}
\begin{aligned}
\ket{\alpha}=\text{e}^{-\left|\alpha\right|^2/2}\sum_n \frac{\alpha^n}{\sqrt{n!}}\ket{n},
\end{aligned} 
\end{equation}
which is a displaced state. Displaced states are the mathematical representation of forms of coherent light, such as laser light. As such, this description is important for applications such as imaging, where a laser is often used as a known source. The statistical properties of this state also serve as a basis for the statistical measures of coherence and therefore will serve as a baseline for control. 

Mathematically, displaced states also provide a useful basis for us to work in. Unlike the fock basis they form an overcomplete basis, mathematically this means that the inner product between two states $\alpha$ and $\beta$ is 
\begin{equation}
\begin{aligned}
\langle\beta|\alpha\rangle=\text{exp}\left[\frac{i}{2}\text{Im}\left(\alpha\beta^*\right)\right]\text{exp}\left[-\frac{1}{2}\left|\beta-\alpha\right|^2\right].
\end{aligned} 
\end{equation} 
Furthermore, the creation of a displaced state is described via the displacement operator
\begin{equation}
 \hat{D}\left(\alpha\right)=\text{exp}\left(\alpha\hat{a}^{\dagger}-\alpha^*\hat{a}\right),
\end{equation}
which acts on the vacuum as $\hat{D}\left(\alpha\right)\ket{0}=\ket{\alpha}$. The displacement operator also acts on a coherent state as
\begin{equation}\label{Chap1:Eqn:CohstateComb}
\begin{aligned}
\hat{D}\left(\beta\right)\ket{\alpha}=\text{exp}\left(i\text{Im}\left[\beta\alpha^*\right]\right)\ket{\alpha+\beta},
\end{aligned}
\end{equation}
allowing for an easy description of coherent states in the vacuum.
\section{P-functions}
\label{Intro:P-functions}
Due to the continuous nature of the displaced states and the ease at which the displacement operator describes the addition of states, representing any state in the coherent basis will be a useful tool. Additionally this should be applicable to mixed as well as pure states. The solution to this problem is the Glauber-Sudarshan P-function \cite{glauber1963quantum}. To begin, consider a state given by the density matrix $\hat{\rho}$ which acts on with the states $\ket{-\alpha}$ and $\ket{\alpha}$ giving \cite{glauber1963quantum}
\begin{equation}
\begin{aligned}
\bra{-\alpha}\hat{\rho}\ket{\alpha}&=\int P\left(\beta\right)\langle-\alpha |\beta\rangle\langle\beta|\alpha\rangle d^2\beta\\
&=e^{-|\alpha|^2}\int P\left(\beta\right)e^{-|\beta|^2}e^{\beta^*\alpha-\alpha^*\beta}d^2\beta.
\end{aligned}
\end{equation}
Which is a 2D Fourier transform over the complex plane. Therefore, solving for the P-function simply by inverting the Fourier transform yields \cite{PhysRevLett.18.752}
\begin{equation}\label{chap1:Eqn:PfuncDef}
\begin{aligned}
P\left(\beta\right)=\frac{e^{|\beta|^2}}{\pi^2}\int e^{|\alpha^2|^2}\bra{-\alpha}\hat{\rho}\ket{\alpha}e^{\alpha^*\beta-\alpha\beta^*}d^2\alpha.
\end{aligned}
\end{equation}
With this we can derive the P-functions for our previously mentioned states. For a coherent states $\ket{\beta}$ the P-function is
\begin{equation}\label{Chap1:Def:CohPfunction}
\begin{aligned}
P\left(\alpha\right)=\delta^2\left(\alpha-\beta\right).
\end{aligned}
\end{equation}
Where $\delta\left(\dots\right)$ is the dirac delta function. For a thermal state with a mean photon number of $\bar{n}$ we can write the P-function as
\begin{equation}\label{Chap1:Def:ThermalPfunction}
\begin{aligned}
P\left(\alpha\right)=\frac{1}{\pi\bar{n}}\text{exp}\left(-\frac{\left|\alpha\right|^2}{\pi\bar{n}}\right).
\end{aligned} 
\end{equation}
The P-function ceases to be have a useful definition when moving to more quantum states, such as the number states we originally derived. At this point the function becomes highly singular. 
It important to note that unlike most density matrix representations where the weight is a probability distribution, that the P-function is a quasiprobability distribution, capable of being negative or highly singular. This comes about through the violation of Kolmorogov's axioms of probability theory, in particular the $\sigma$-additivity axiom, where probabilities must yield the result of mutually exclusive states \cite{doi:10.1063/1.522736,borovkov1999probability}. These traits appear when the state of light starts exhibit non-classical properties such as entanglement and superposition, which allow for states to have properties different from standard classical fields. As such, these features serve as an indicator of quantum forms of light, and as I'll discuss in section \ref{QCoh} will have properties not predicted by classical theories.

\section{Polarization of Light}
Polarization of photons is an essential quantity to quantum technologies \cite{bib:BennetBrassard84,bib:Pittman01}. In Chapter \ref{Quantum Networks} the qubit will be encoded into the polarization \cite{bib:riedl2012bose}, as well as in Chapter \ref{MQvCZT} where the coherence of various post-selected polarization measurements will yield the main result \cite{PhysRevA.93.053836}. As such, a proper knowledge of polarization is essential for this paper. Polarization consists of two mutually exclusive polarization states. Conventionally, there are three important sets that are used, the horizontal/vertical $(\ket{H},\ket{V})$, the diagonal/anti-diagonal $(\ket{A},\ket{D})$ and the left/right circularly polarized $(\ket{L},\ket{R})$, These three basis are not orthogonal from each other thus the state of one basis can be written as a superposition of two states in another basis. For example, the diagonal state can be written as $\ket{D}=\frac{1}{\sqrt{2}}\left(\ket{H}+\ket{V}\right)$ \cite{PhysRevA.93.053836,bib:BennetBrassard84}. This creates an interesting property of quantum pictures of polarization. To emphasize this point, consider the projection operator $\hat{P}_i=\ket{i}\bra{i}$, where $i$ represents the polarization of the projection which represents a polarizer. Now, consider a case where a diagonal source of light is passed through a horizontal polarizer, and then through an anti-diagonal polarizer. In this case, there is a $1/4$ chance that the original photon is detected on the other side of the anti-diagonal polarizer. Comparatively the same system with the horizontal polarizer will never have a photon detected after the anti-diagonal polarizer. Allowing multiple projective measurements to effectively rotated the polarization of a photon, forming an essential property of polarization that will be exploited in Chapter \ref{MQvCZT}.

The polarization properties of photons form a two-dimensional Hilbert space, over which any set of two mutually exclusive states can be declared, with a conjugate pair, forming a mutually unbiased basis \cite{magana2019quantum}. This means that the repeated projection of one member of the mutually unbiased basis into another leads to an equal probability of measurement of all elements in the projecting basis. As such the outcome of projection measurements is completely random and unbiased. This is an essential feature for quantum number generation and cryptography protocols such as BB84 \cite{bib:BennetBrassard84}.
\section{The Rayleigh-Abbe Criterion}
One of the major classical limits in imaging is the Rayleigh-Abbe criterion \cite{rayleigh1879xxxi}. This criterion creates a diffraction limit through which finer detail on an object cannot be resolved. In the case of viewing point sources, this limit sets the minimum distance through which two point sources can be resolved. The Rayleigh criterion can be found by considering the point spread function of two point sources of equal strength. I assumed that each source has the point spread function of an Airy disk, which is described by the first Bessel function. Thus, the angular resolution can be found to be
\begin{align}
 \theta=1.22 \frac{\lambda}{D}, 
\end{align}
where $\theta$ is the angular resolution, $\lambda$ is the wavelength of the source and $D$ is the diameter of the lens. This can be used to find a spatial resolution of we can take the small angle approximation and find
\begin{align}
 \Delta l\approx 1.22 \frac{f\lambda}{D}, 
\end{align}
 where f is the focal length of the objective lens. Describing the spatial resolution by the numerical apeture instead gives a limit of
\begin{align}
 \Delta l = .61 \frac{\lambda}{\text{NA}}. 
\end{align}
where NA is the numeric apeture. 

The Abbe criterion provides an even smaller minimum resolvable here Abbe defined the resolvable limit as
\begin{align}
 \Delta l=\frac{\lambda}{2\text{NA}}. 
\end{align}
 This largely comes from the fact that Abbe defined used a different definition for resolvable. In the case of the Rayleigh limit the first minima of the Airy function overlaps the maxima of the second Airy function. Abbe instead insisted that the beams could be closer, defining a distance where any local minima exists between the two maxima as resolvable. 

\section{Feed Forward Neural Networks}
Another important tool for developing quantum technologies is machine learning \cite{you2020identification,Perrier_2020,bhusal2021spatial,lollie2021highdimensional,bhusal2021smart,krenn_computer-inspired_2020}, which I will cover in the form of feed forward neural networks \cite{bebis1994feed}. The goal of any machine learning algorithm is to take some number of observations and convert it into a prediction. There exist three main forms of machine learning supervised learning \cite{bebis1994feed}, unsupervised learning \cite{knnOriginal} and reinforcement learning \cite{kybernetes:forsyth}. In the case of supervised learning an algorithm is created and then fed a series of data with the correct labels. It then adjusts weights within the network in order to receive data and accurately find its label. Unsupervised learning isn't given data with labels, as such its main objective is to find some sort of pattern in a set of unlabeled data. Reinforcement learning works by interacting with an environment where its actions are judged by a reward function. The goal of the algorithm is thus to find the optimal means of interacting judged by a maximum value of its rewards function.

For this dissertation the main concern will be supervised learning. In Chapter \ref{QI}, a feed forward network will be used in order to allow imaging beyond the Rayleigh-Abbe criterion 
\cite{bhusal2021smart}. To introduce the mechanism through which a feed forward network function I will first introduce its building block, the neuron.

Neuron's are simple mathematical constructions that consist of three components; weighted inputs, an activation function, and an output \cite{aggarwal2018neural}. A single neuron allows for simple binary identification. To train, a set of data is fed through the weighted inputs to the activation function, which in turn outputs a value as expected. This value can then be interpreted as one of two classes. When training the output is identified as either one of the two classes, in the cases of a misidentification, the weights are adjusted. Throughout many iterations this will allow the weights to become properly adjusted for optimal classification.

The feed forward network is the multi-neuron generalization of a single neuron \cite{aggarwal2018neural,bebis1994feed}. Here, there are at least three neuron layers, each consisting of some number of neurons. However, the final neuron layer must have as many neurons as classes. Each layer of neurons is fully connected to the previous layer in the initial network. The first layer is the input layer, it contains as many neurons as inputs, this layer simply serves to take data and pass it to the first hidden layer. The hidden layers, of which there could be one or more, perform the majority of the data processing. Each neuron takes the data provided by its weights and performs a transformation to it, before passing it on to the next layer. Once all of the hidden layers are process the data, the final result is passed to the output layer, which puts the data into a form to be classified. Once the neural network is designed training begins. For training a network must have three pieces defined, a cost function, a loss function and a gradient algorithm. The cost function defines the current error in classification, and should be defined such that the best classification has the minimum cost. This however is defined as a function of weights in the neural network. The loss function is similar to the cost function, however it compares the true and predicted values in order to determine the functional accuracy of a given set. Finally, the gradient algorithm seeks to optimize the weights in a graph as to minimize error. As such it will take the cost and loss functions and seek to adjust weights to minimize loss \cite{aggarwal2018neural}. This requires a large set of training data since the using the whole set in each iteration can leave to issues such as over fitting.

The central problem with neural network design comes in the form of two problems, underfitting and overfitting. Underfitting as its name implies, fails to create a neural network capable of learning all of the features of a given data set. This of course can lead to high misidentification rates. Overfitting on the other hand, learns too many features of given data set, mistaking the noise for a standard feature \cite{aggarwal2018neural}. While this may allow for proper identification for the training sets, it often leads to the failure to perform otherwise. As such, a testing step is with a known data set other than the training set must be performed to ensure proper performance.
 \singlespacing
\chapter*{Chapter 2. \\ Coherence}
\addcontentsline{toc}{chapter}{Chapter 2. Coherence}
\setcounter{section}{0}
\setcounter{figure}{0}
\setcounter{table}{0}
\refstepcounter{chapter}
\label{Coherence}
\doublespacing
The practical implementation of light requires a model on how light interferes and interacts.  Furthermore, this model must be able to extend to higher orders and be quantifiable, allowing for the creation of a quantum picture.  Throughout this chapter a basis in coherence, both classical and quantum will be established, demonstrating the mathematical tools and theorems which in aid in the creation of new quantum technologies.

This chapter begins with a discussion of classical coherence covering the Young's dual slit experiment.  From there the van Cittert-Zernike theorem and the beam coherence polarization matrix will be covered in detail, which will be essential to the content of Chapter \ref{MQvCZT}.  The Hanbury Brown Twiss experiment will be covered allowing for the demonstration of a quantum picture of coherence \cite{HBTExp}.  

\section{Spatial Coherence}
\label{ClassCoh}
\begin{figure}
\centering
\includegraphics[scale=0.2]{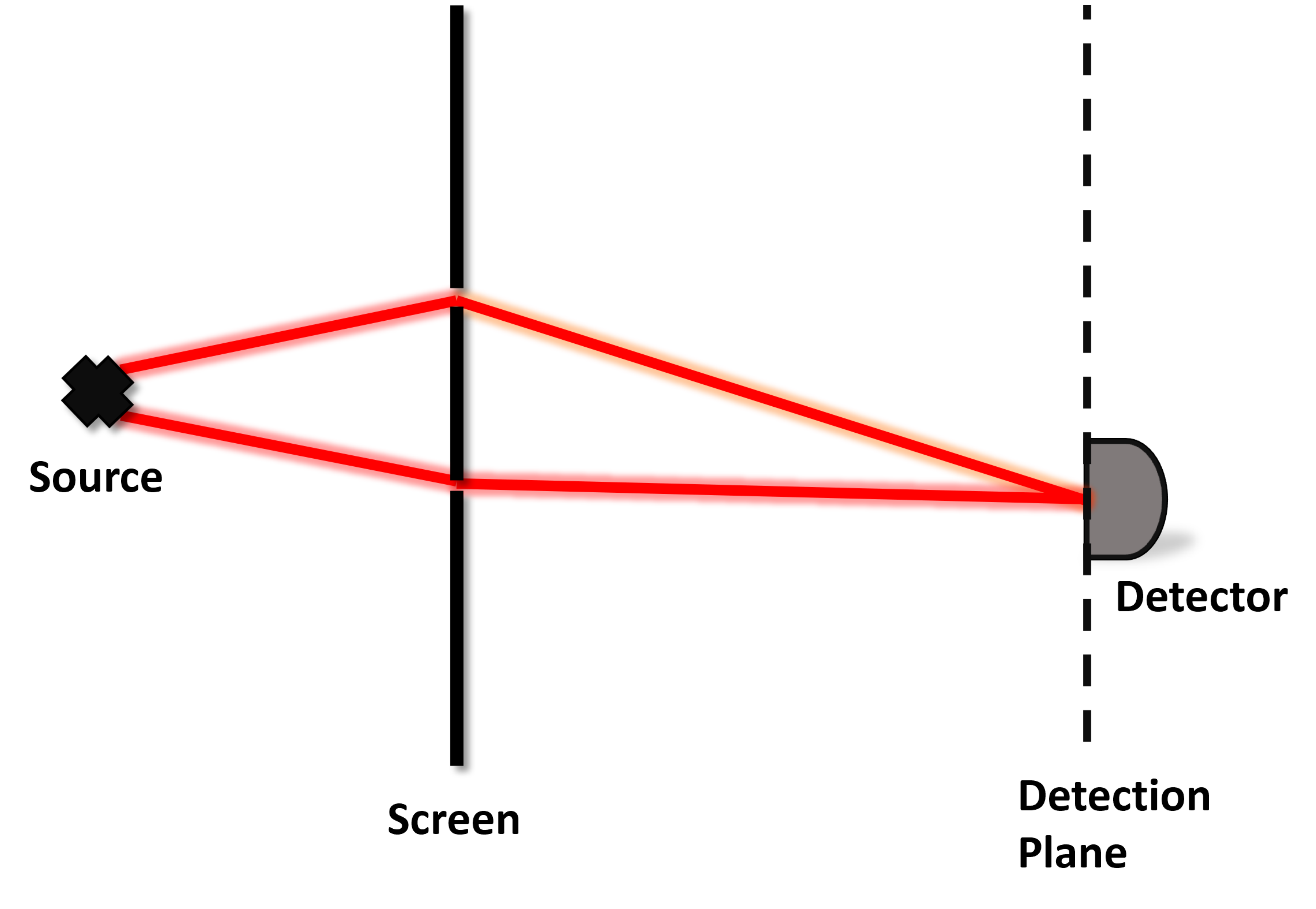}
\caption[A diagram showing Young's dual slit experiment.]{A diagram showing Young's dual slit experiment. Each slit is at the position $x=\pm d/2$ on the screen. Light then propagates to the screen where the first-order correlation is measured.}
\label{fig:DualSlit}
\end{figure}
A simple model of first order coherence can be observed though the analysis of Young's dual slit experiment shown in Figure \ref{fig:DualSlit}. To begin, consider a monochromatic horizontally polarized thermal source incident on a screen at $z=0$ \cite{Rodenburg:14,gerry2005introductory,mandel1995optical}.  The screen has two slits at a distance $d$ apart and are of a width that we can ignore diffraction effects. Since we are assuming that the two fields have the same polarization we can treat them as scalar fields, resulting in the field immediately after the screen to be
\begin{equation}
\begin{aligned}
E\left(\mathbf{r},t\right)=K_1 E\left(\mathbf{r}_1,t_1\right)+K_2 E\left(\mathbf{r}_2,t_2\right).    
\end{aligned}
\end{equation}
Where $K_1$ and $K_2$ are complex amplitudes that are related to the path length. This term holds for any point after the screen with differences in position accounted for by both $K_i$ and $E(\dots)$. At the observation plane, interference fringes emerge which are a manifestation of spatial coherence \cite{PhysRevLett.98.043908,PhysRevLett.122.133601,Gbur:03,magana-loaiza_exotic_2016}. The spatial coherence is caused by the introduction of the spatial seperation between the two slits.  Mathematically, the interference pattern can be found by solving for the intensity which is 
\begin{equation}\label{Chap2:Eqn:Coh2}
\begin{aligned}
I\left(\mathbf{r}\right)&=\left\langle\left|E\left(\mathbf{r},t\right)\right|^2\right\rangle\\
&=\left|K_1\right|^2\left\langle\left|E\left(\mathbf{r}_1,t_1\right)\right|^2\right\rangle+\left|K_2\right|^2\left\langle\left|E\left(\mathbf{r}_2,t_2\right)\right|^2\right\rangle\\
&+2\text{Re}\left[K_1^*K_2\left\langle E^*\left(\mathbf{r}_1,t_1\right) E\left(\mathbf{r}_2,t_2\right)\right\rangle\right].
\end{aligned}    
\end{equation}
The equation above  consists of three terms, the first two are the intensities related to the left and right slit alone and the final terms is related to the spatial coherence generated between the two slits. It is important to note that interference fringes are only generated between the two sources if $\Delta\theta d/2\leq\lambda$ where $\Delta\theta$ is the distance of the separation between the two pinholes subtends at the source and $\lambda$ is the wavelength of the beam.  The interference is only generated in an area $\Delta A=4 R^2 \lambda^2/d^2$ where R is the distance between the beam source and the screen.  This area is known as the coherence area and its square root is the coherence length.

The third term of Eqn. (\ref{Chap2:Eqn:Coh2}) enables a description of first-order coherence.  Normalizing this term defines the classical first order coherence function to be \cite{Mirhosseini2016}
\begin{equation}\label{Chap2:Def:1st order coherence}
\begin{aligned}
\gamma^{(1)}(x_1,x_2)=\frac{\left\langle E^*(x_1)E(x_2)\right\rangle}{\sqrt{\left\langle\left|E(x_1)\right|^2\right\rangle\left\langle\left|E(x_2)\right|^2\right\rangle}}.
\end{aligned}    
\end{equation}
Where $x_i$ simply refers to the electric field at points $x_1$ and $x_2$ on the detection screen. The above formula can be used to describe the coherence a beam of light is at any point.  A value of one indicates complete coherence, less than one is partial coherence and zero is completely incoherent. This quantity proves vital when  establishing the van Cittert-Zernike theorem in the next section. It is also important to note that a value of one does not always correspond to a truly coherent source since there are many degrees of coherence that contribute to the coherence of a single source.  While  a description of these higher-order correlations could be created now only using the classical fields, such a description would be quickly discarded for this dissertation.  As I will discuss in Section \ref{QCoh} starting at second-order the quantum and classical coherence functions will have different properties, causing only the use of the quantum second-order functions for this disseration.

\section{Van Cittert-Zernike Theorem}
\label{vCTZ}

One of the main theorems for partial coherence is the van-Cittert Zernike theorem \cite{ZERNIKE:38,Cittert:34,mandel1995optical}.  This theorem provides the formalism to understand how coherence properties change as the field propagates. To begin, consider a complex field at two points i.e. $V(\mathbf{r}_1',t)$ and $V(\mathbf{r}_2',t)$ where $\Tilde{V}(\mathbf{r}_i,t)$ is the spectral amplitude of the field.  We assume the fields are stationary and ergodic.  The fields at the points of measurement can then be found by the Huygens-Fresnel principle to be \cite{goodman2005introduction}
\begin{equation}\label{Chap:Eqn:Fields}
\begin{aligned}
\Tilde{V}\left(\mathbf{r}_i,\nu\right)=\int_\mathcal{A}\Tilde{V}\left(\mathbf{r}'_i,\nu\right)\frac{e^{ikR_i}}{R_i}\Lambda_i d^2r'_i.
\end{aligned}   
\end{equation}
Where the subscript $i$ denotes which field, $\Lambda_i$ is the inclination factor, $R_i$ is the distance from a point on the surface to the a point on the propagation plane and $k$ is the wavenumber. Multiplying the complex conjugate of field 1 with field 2 and taking the ensemble average gives the cross-spectral density function propagation law of
\begin{equation}
\begin{aligned}
W\left(\mathbf{r}_1,\mathbf{r}_2,\nu\right)=\int_\mathcal{A}\int_\mathcal{A}W\left(\mathbf{r}_1'\mathbf{r}_2',\nu\right)\frac{e^{ik\left(R_2-R_1\right)}}{R_1R_2}\Lambda_1^*\left(k\right)\Lambda_2\left(k\right)d^2r_1' d^2r_2'.
\end{aligned}    
\end{equation}
Assuming the light is quasi-monochromatic with a mean frequency of $\bar{\nu}$ a formula can be derived describing the propagation of the mutual coherence.  Neglecting the dependance of $\Lambda_i$ on frequency, instead describing it as a mean value $\bar{\Lambda}_i$. Then multipling both sides by $e^{-2\pi i\nu\tau}$ and integrating over frequency. Finally, the mutual coherence function when the retardation time is small compared to the coherence time is
\begin{equation}\label{Chap2:Eqn:MutCoh}
\begin{aligned}
\Gamma\left(\mathbf{r}_1,\mathbf{r}_2,\tau\right)\approx\int_\mathcal{A}\int_\mathcal{A}\Gamma\left(\mathbf{r}'_1,\mathbf{r}_2',\tau\right)\frac{e^{i\bar{k}\left(R_2-R_1\right)}}{R_1R_2}\bar{\Lambda}_1^*\bar{\Lambda}_2d^2r_1'd^2r_2'.
\end{aligned}    
\end{equation}
 
Now Zernike's propagation law can be derived with very little effort.  Starting with Eqn. (\ref{Chap2:Eqn:MutCoh}) I can let $\tau=0$ and restrict the system using the small angle propagation.  This returns Zernike's propagtion law as
\begin{equation}\label{Chap2:Def:ZernikesPropLaw}
\begin{aligned}
 J\left(\mathbf{r}_1,\mathbf{r}_2\right)=\left(\frac{\bar{k}}{2\pi}\right)^2\int_\mathcal{A}\int_\mathcal{A}J\left(\mathbf{r}_1',\mathbf{r}_2'\right)\frac{e^{i\bar{k}\left(R_2-R_1\right)}}{R_1R_2}d^2r_1'd^2r_2'.
\end{aligned}    
\end{equation}
Where $J\left(\mathbf{r}_1,\mathbf{r}_2\right)$ is the mutual intensity.  

Now letting the surface $\mathcal{A}$ coincide with a radiating surface $\sigma$ of a spatially incoherent, planar, quasi-monochromatic source.  Thus the initial mutual coherence function on the surface is $J\left(\mathbf{r}_1',\mathbf{r}_2\right)=I\left(\mathbf{r}_1'\right)\delta^{(2)}\left(\mathbf{r}_2'-\mathbf{r}_1'\right)$. Where $I\left(\mathbf{r}\right)$ is the intensity at $\mathbf{r}$ and $\delta^{(2)}$ is the two-dimensional Dirac delta function.  Substituting the initial intensity into Zernike's propagation law and normalizing recovering the van Cittert-Zernike theorem as
\begin{equation}\label{Chap2:Def:vCZT}
\begin{aligned}
\gamma^{(1)}\left(\mathbf{r}_1,\mathbf{r}_2\right)=\frac{1}{\sqrt{I\left(\mathbf{r}_1\right)I\left(\mathbf{r}_2\right)}}\left(\frac{\bar{k}}{2\pi}\right)^2\int_\sigma I\left(\mathbf{r}'\right)\frac{e^{i\bar{k}\left(R_2-R_1\right)}}{R_1R_2}d^2r'.
\end{aligned}    
\end{equation}
Where $\gamma^{(1)}$ is the spatial coherence we define in Eqn. (\ref{Chap2:Def:1st order coherence}).  Therefore, the van Cittert-Zernike theorem defines the propagation of first order coherence.

\section{Beam Coherence Polarization Matrix}
\label{BCP Matrix}

In the study of coherence it is also important to relate the spatial coherence to polarization for a full understanding of unpolarized and multimode fields.  This formalism is provided by the beam coherence polarization (BCP) matrix \cite{gori1998beam}.  The BCP matrix measures the first-order coherence between various polarization components. More importantly, the BCP matrix can be easily propagated using beam like propagators, making it an ideal candidate for modeling the change in coherence as the field propagates.

In order to derive the BCP matrix, consider that most partially polarized fields are beam shaped and therefore closely gathered around the center. Then introducing the covariance matrix of the electric field as
\begin{equation}\label{Chap2:Eqn:BCP1}
\begin{aligned}
\Gamma\left(\mathbf{r}_1,\mathbf{r}_2,z;\tau\right)=\begin{bmatrix}
\Gamma_{xx}\left(\mathbf{r}_1,\mathbf{r}_2,z;\tau\right) && \Gamma_{xy}\left(\mathbf{r}_1,\mathbf{r}_2,z;\tau\right)\\
\Gamma_{yx}\left(\mathbf{r}_1,\mathbf{r}_2,z;\tau\right) && \Gamma_{yy}\left(\mathbf{r}_1,\mathbf{r}_2,z;\tau\right),
\end{bmatrix}
\end{aligned}    
\end{equation}
where each element is defined by $\Gamma_{\alpha\beta}=\left\langle E_{\alpha}^*\left(\mathbf{r}_1,z,t\right)E_{\beta}\left(\mathbf{r}_2,z,t+\tau\right)\right\rangle$. Where $t$ is the time variable, $\tau$ is the delay between the two fields, $E_\alpha$ is the $\alpha$ polarized component of the electric field, the asterisk denotes the complex conjugate and the angle brackets denote time average.  Note that the field is monochromatic with a mean frequency of $\bar{\nu}$ and an effective bandwidth of $\Delta\nu$. Now by letting the time delays from the apeture be small compared to the coherence time $1/\Delta\nu$ we find an individual element of the field to be
\begin{equation}\label{Chap2:Eqn:Approx}
\begin{aligned}
\Gamma_{\alpha\beta}\left(\mathbf{r}_1,\mathbf{r}_2,z;\tau\right)=J_{\alpha\beta}\left(\mathbf{r}_1,\mathbf{r}_2,z\right)\text{exp}\left(-2\pi i\bar{\nu}\tau\right).
\end{aligned}    
\end{equation}
Where  $J_{\alpha\beta}\left(\mathbf{r}_1,\mathbf{r}_2,z\right)=\Gamma_{\alpha\beta}\left(\mathbf{r}_1,\mathbf{r}_2,z;0\right)$, which denotes the coherence and polarization of the field for a certain measurement.  Proceeding, the BCP matrix can now be written as
\begin{equation}\label{Chap2:Eqn:BCPMatrix}
\begin{aligned}
\hat{J}(\mathbf{r}_1,\mathbf{r}_2,z)=\begin{bmatrix}
J_{xx}\left(\mathbf{r}_1,\mathbf{r}_2,z\right) && J_{xy}\left(\mathbf{r}_1,\mathbf{r}_2,z\right)\\
J_{yx}\left(\mathbf{r}_1,\mathbf{r}_2,z\right) && J_{yy}\left(\mathbf{r}_1,\mathbf{r}_2,z\right)
\end{bmatrix}.
\end{aligned}
\end{equation}
The beam polarization matrix provides a useful formalism for propagating the coherence and polarization throughout the system.  The intensity at any point $\mathbf{r}$ is found by $I(\mathbf{r},z)=\text{Tr}\left[\hat{J}\left(\mathbf{r},\mathbf{r},z\right)\right]$.  The BCP matrix can be normalized, where it is then described as \cite{Mirhosseini2016}
\begin{equation}\label{Chap2:Def:BCPNorm}
\begin{aligned}
j_{\alpha\beta}\left(\mathbf{r}_1,\mathbf{r}_2,z\right)=\frac{J_{\alpha\beta}\left(\mathbf{r}_1,\mathbf{r}_2,z\right)}{\sqrt{J_{\alpha\alpha}\left(\mathbf{r}_1,\mathbf{r_1},z\right)J_{\beta\beta}\left(\mathbf{r}_2,\mathbf{r}_2,z\right)}}.
\end{aligned}
\end{equation}
Where each element is now the first order coherence $\gamma^{(1)}$ in Eqn. (\ref{Chap2:Def:1st order coherence}) for that polarization measurement. Furthermore, the local degree of polarization can be found by
\begin{equation}\label{Chap2:Def:Polarization}
\begin{aligned}
P\left(\mathbf{r},z\right)=\sqrt{1-\frac{4\text{det}\left[\hat{J}\left(\mathbf{r},\mathbf{r},z\right)\right]}{\left(\text{Tr}\left[\hat{J}\left(\mathbf{r},\mathbf{r},z\right)\right]\right)^2}}
\end{aligned}    
\end{equation}
The biggest advantage to the BCP matrix is its ability to be propagated through polarizing material.  Consider an optical element described by the jones matrix $\hat{T}\left(\mathbf{r}\right)$.  The BCP matrix upon propagating though the grating is given as
\begin{equation}\label{Chap2:Eqn:BCPPol}
\begin{aligned}
\hat{J}'\left(\mathbf{r}_1,\mathbf{r}_2,z\right)=\hat{T}\left(\mathbf{r}_1\right)\hat{J}\left(\mathbf{r}_1,\mathbf{r}_2,z\right)\hat{T}\left(\mathbf{r}_2\right).
\end{aligned}    
\end{equation}
Further propagation of the BCP matrix through free space can be done following the formalism set in Section \ref{vCTZ}, since the BCP matrix consists of coherence elements.

\section{Quantum Coherence}
\label{QCoh}
\begin{figure}
\centering
\includegraphics[scale=0.2]{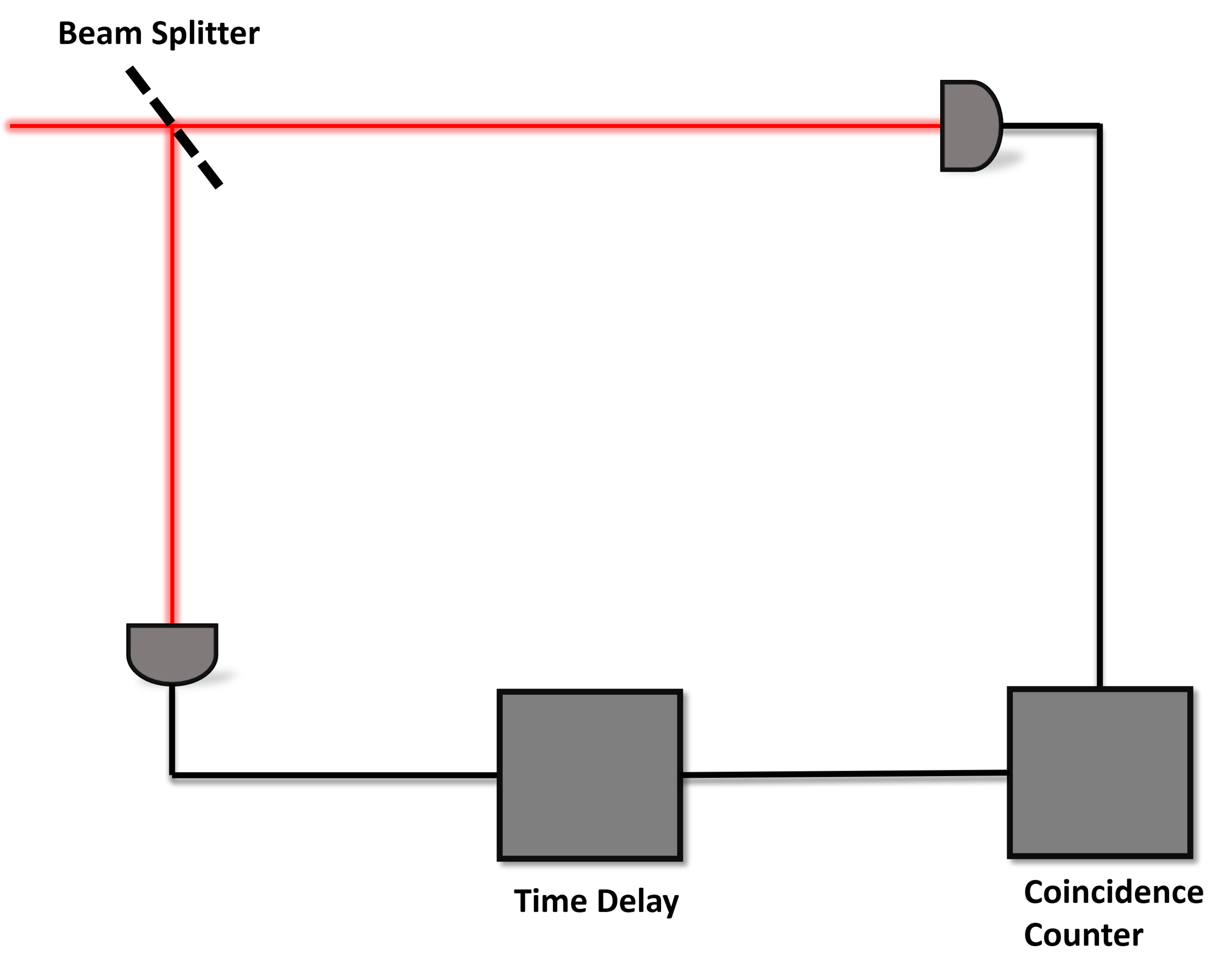}
\caption[A diagram showing of the Hanbury Brown Twiss experiment.]{A diagram showing of the Hanbury Brown Twiss experiment.  Here a thermal light beam is incident on a fifty-fifty beam spitter, where it is split into two equal intensity beams.  The beams travel an equal distance before being sent to a detector.  One of the two paths has a time delay set, allowing the time correlations throughout the experiment to be measured. Once both signals are measured their coincident counts are found.}
\label{fig:HBT}
\end{figure}
In the 1950's Hanbury, Brown and Twiss developed a temporal correlation experiment that measured the correlations between intensities rather than the correlations between fields as discussed so far \cite{HBTExp,brown_correlation_1956}.  In this experiment a single source of light is passed through a beam splitter and sent down two paths.  At an equidistant end of each path sat a detector capable of measuring the intensity of light.  One detector  had a variable time delay placed in its arm set to time $\tau$, so that if detector one took a measurement at time $t$, detector two would take a measurement at time $t+\tau$.  In the case that $\tau$ is less than the coherence time of the beam information on the photon statistics of the beam could be measured.  

Originally, Hanbury Brown and Twiss argued the results by assuming that photons were emitted independently by the source \cite{CoherenceReview}.  Therefore, by assuming that photons would be not be split by the beam splitter but rather only be transmitted or reflected.  However, what they
 found was that for zero time delays the measurement gave the same result as for time delays.  This indicates that there is another effect, known as bunching was occuring. In the case of bunching photon do not travel independently from one another but rather in pairs.  This result gave birth to the field of quantum optics and was later explained by the quantum theory of light described by Glauber \cite{glauber1963quantum}.  However, as we will show later in this chapter, there exists inter-relating photon effects similar to bunching that can only be found by the quantum picture, establishing the need to expand our current definition of coherence.
 
 To begin building a quantum picture of coherence I start with the first order as we have covered so far.  The quantum first-order of coherence is defined as
\begin{equation}
\begin{aligned}
G^{(1)}\left(x_1,x_2\right)=\text{Tr}\left[\hat{\rho}\hat{E}^{(-)}\hat{E}^{(+)}\right].
\end{aligned}     
\end{equation}
Which is normalized similarly to Eqn. (\ref{Chap2:Def:1st order coherence}) by just replacing the classical fields with the field operators and finding the ensemble average. The intensity correlation of the field can be found using the equation
\begin{equation}
\begin{aligned}
G^{(2)}\left(x_1,x_2;x_1,x_2\right)=\text{Tr}\left[\hat{\rho}\hat{E}^{(-)}(x_1)\hat{E}^{(-)}(x_2)\hat{E}^{(+)}(x_2)\hat{E}^{(+)}(x_1)\right]
\end{aligned}    
\end{equation}
Which can be reformulated in terms of delays and normalized giving the second-order coherence function to be
\begin{equation}\label{Chap2:Def:g2}
\begin{aligned}
g^{(2)}\left(\tau\right)=\frac{\left\langle\hat{E}^{(-)}(t)\hat{E}^{(-)}(t+\tau)\hat{E}^{(+)}(t+\tau)\hat{E}^{(+)}(t)\right\rangle}{\left\langle\hat{E}^{(-)}(t+\tau)\hat{E}^{(+)}(t+\tau)\right\rangle\left\langle\hat{E}^{(-)}(t)\hat{E}^{(+)}(t)\right\rangle}
\end{aligned}    
\end{equation}
Which can be measured either by analyzing photon statistics or by finding time correlations.  

Quantum second-order coherence is essential for understanding the properties of light sources. For coherent light  $g^{(2)}\left(0\right)=1$, and for thermal light $g^{(2)}\left(0\right)=2$.  This provides rudimentary identification metric that will be applied in detail in Chapter \ref{QI}. Furthermore, these values can extend from $[0,\infty)$.  For a $g^{2}\left(0\right)<1$ we can note that our state has sub-Poissonian statistics, which indicates it can be used for applications such as sub-shot noise sensing.  Even more interesting if $g^{2}\left(\tau\right)>g^{2}\left(0\right)$ an  effect known as anti-bunching occurs, in which  photons arrive evenly spaced in time.  Using metrics such as the Glauber-Sudarshan P-function it becomes clear that these states are highly non-classical and possess a great degree of correlations unpredictable by classical theories, as discussed in Section \ref{Intro:P-functions} \cite{glauber1963quantum}.  The $g^{(2)}$ function will bring to light such states in Chapter \ref{MQvCZT}.

 \singlespacing
\chapter*{Chapter 3. \\ Symmetries in Quantum Networks}
\addcontentsline{toc}{chapter}{Chapter 3. Symmetries in Quantum Networks}
\setcounter{section}{0}
\setcounter{figure}{0}
\setcounter{table}{0}
\refstepcounter{chapter}
\label{Quantum Networks}
\doublespacing
The central technology of the information age is the internet.  From this humans are connected more than ever, allowing us to communicate, share information, and access personal records with more efficiency and ease than ever before in human history.  This modern marvel depends on direct communication that transmits classical information.  However, as our knowledge of quantum technologies advances the failings of the internet as we know it begin to show.   The potential of quantum computers poses an existential threat to the classical internet due to their potential ability to easily crack RSA encryption \cite{bib:shor1994algorithms,bib:DeutschJozsa92,bib:Deutsch85}.  Additionally, our current infrastructure can only relays classical information, preventing distributed computing with quantum computers \cite{beals2013efficient,cacciapuoti2019quantum}.

These failures of classical networking necessitate an advancement worthy of other quantum technologies in development, which leads to the creation of the quantum network.  Quantum networks, as their name implies, transmit quantum information rather than classical information.  This can be achieved through entangle states, though other means exist as well \cite{bhusal2021spatial,lollie2021highdimensional,Mirhosseini_2015,bib:BennetBrassard84,bib:Ekert91}. The importance of entangled states is that they allow the circumvention of the no cloning theorem, which prevents the direct copying of unknown quantum states \cite{bib:Bennett1993,gerry2005introductory}.  Entangled states however, can communicate arbitrary quantum information using repeat-until-success entanglement transmission and quantum teleportation \cite{bib:PRL_70_1895,bib:Boumeester97,pirandola_advances_2015}. Combined with entanglement swapping and purification we are able to build a quantum network of an arbitrary size \cite{rohde2021quantum,bib:PRL_80_3891,PhysRevA.60.194,PhysRevA.57.822,bib:Pan01}.  Such a network is essential for facilitating future quantum technologies, such as quantum key distribution, quantum state teleportation, and distributed quantum computing \cite{beals2013efficient,bib:BennetBrassard84,bib:PRL_68_557,bib:Bennett1993,cacciapuoti2019quantum,bib:NielsenChuang00}.

In this chapter, I will create a set of rules that should guide future work in quantum network routing.  To do so I will introduce a simple dephasing and loss model over each channel of the quantum network.  Then, I will use the operation of entanglement swapping and purification to build an algebra capable of performing graph reductions. This algebra will help create a cost vector, defining quantities that should be minimized and maximized in order to have optimal lossless transmission.  Lastly, I will propose a rudimentary algorithm capable of finding the optimal sub graph for transmission.  This chapter represents my contribution to Leone et al \cite{leone2021qunet}.

\section{Entanglement Distribution Networks}
\label{Network Symmetries}
My central goal is determine the optimal "path" through a network. As such I am aiming to exploit the quantum network to produce a minimal amount of error. Classically, the network would be represented by a graph where each edge is weighted with a loss  and the optimal path would be the shortest route between two points.  However, since quantum networks are prone to more error than simple loss a more complicated model of error must be created which accounts for a wide variety of errors \cite{Proctor2018_PRL,Sidhu2020_AVS}. To begin I represent the network using the graph $G=(V,E).$ Where the graph vertices $V$ represent nodes, and define end-users, or devices that implement quantum operations (such as entanglement purification) \cite{bib:DurBriegel999,PhysRevA.60.194}. Alternatively, a node can also act as a router to conduit entanglement between other nodes in the graph through entanglement swapping \cite{bib:jennewein2001experimental}. Edges, $E$, between vertices represent lossy quantum channels.

The main resource I aim to exploit in a quantum network is entanglement. For the network, which I assume relies on optical qubits encoded using polarization \cite{gerry2005introductory}, this resource is provided by Bell states. There are four Bell states, all locally equivalent up to Pauli $\hat{X}$ and $\hat{Z}$ operations, which are maximally entangled and defined through \cite{braunstein1992maximal}
\begin{align}
    \ket{\Phi^\pm}_{A,B} &= \frac{1}{\sqrt{2}} (\ket{0}_A\ket{0}_B \pm \ket{1}_A\ket{1}_B)\\
    \ket{\Psi^\pm}_{A,B} &= \frac{1}{\sqrt{2}} (\ket{0}_A\ket{1}_B \pm \ket{1}_A\ket{0}_B).\nonumber
\end{align}

As my central resource of interest they are subject to the no cloning limitation. However, Bell states have a useful property that Pauli errors commute across the qubits on an entangled pair.  This means that error on one qubit can be taken to mean an error on the other qubit, up to some global phase \cite{bib:NielsenChuang00}. 

I assume the bell states are transmitted through noisy quantum channels. I will define a cost vector consisting of the two dominant forms of noise in real-world implementations: dephasing and loss \cite{bib:RohdeRalph06}. While not performed here, the cost-vector approach can be extended to other forms of noise as well as other costs, such monetary cost.

The quantum error channels under a decohering model take the form
\begin{align}\label{Chap3:Def:GenNoise}
    \mathcal{E}(\hat\rho) = p\hat\rho + (1-p)\hat\rho_{ss}. 
\end{align}
Where $\hat{\rho}$ is the state, $p$ is the success probability and $\hat{\rho}_{ss}$ is the steady state of the channel meaning $\mathcal{E}(\hat\rho_ss)=\hat\rho_{ss}$.  This provides a simple model where transmission succeeds, represented by the first component, or fails in which case the state is replaced with the steady state of the channel.  A state traveling through multiple channels can be described as
\begin{equation}\label{Chap3:Def:LossOvertraversal}
\begin{aligned}
\mathcal{E}_n\circ\dots\circ\mathcal{E}_1(\hat\rho) = \prod_{i=1}^n p_i\hat\rho + (1-\prod_{i=1}^n p_i)\hat\rho_{ss}. 
\end{aligned} 
\end{equation}
Now, using  these two formulas both to expand upon to  additional forms of noise. First, consider a loss channel that has some chance of loss $\eta=1-p$ and upon loss,  the steady state is the quantum vacuum, completely annihilating the qubit.  This allows for coverage of any form of noise that results in the qubit to be lost or the information originally encoded in the qubit to unrecoverable .  The next case of interest is decoherence channels.  For simplicity, the model is restricted to dephasing. Maintaining the form of error in Eqn. (\ref{Chap3:Def:GenNoise}) noting that in the case of dephasing the steady state is \cite{bib:NielsenChuang00}
\begin{align}
\hat\rho_{ss,\text{dephasing}}=\frac{1}{2}\left(\hat\rho+\hat{Z}\hat\rho\hat{Z}\right).
\end{align}
A figure of interest in quantifying the quality of communicated quantum states is their fidelity, which  is simply the overlap between the expected and actual states. Fidelity is defined as \cite{PhysRevLett.113.190404}
\begin{align}
    F &= \bra{\psi}\mathcal{E}(\ket{\psi}\bra{\psi})\ket{\psi}.
\end{align}
It is useful to only consider the fidelity and loss as two separate phenomena.  As such,  a cost-vector is defined as $(F,\eta)$ where $\eta$ is the chance of a successful transmission. By exploring the algebra's of these two metrics with the above operations I will be able to create a set of rules to find the optimal method for quantum communication.
\section{Network Reduction and Optimal Paths}
\label{Costs}
\noindent
In order to create a quantum equivalent of cost-vector analysis, a mathematical formalism must be derived to describe loss during communication.  As stated before there are two modes of transmission, entanglement swapping and entanglement purification. 

\subsection{Entanglement Swapping}

The simpler of these two cases is entanglement swapping, where the successive error of swapping is defined by Eqn. (\ref{Chap3:Def:LossOvertraversal}).  This can easily be converted into a logarithmic form $-\text{log}(p)$ which acts as an additive loss over transmission. It is important to note that this property holds for Bell state networks due to the commutative nature of error.  Furthermore, the actual logistics of distributing bell pairs are neglected by this metric, instead a simplified model where an imperfect distribution is simply accounted for in the error $p$, allowing an easier analysis of the graph properties to be found \cite{bib:OE_13_202}. This is motivated by a desire to show basic properties of a quantum network and establish an early formalism for cost-vector analysis to be performed. This also allows our analysis to be platform independent, which proves advantageous at this current in development.

There are both stochastic and deterministic applications of entanglement swapping, with stochastic implementations being the main application in a photonic context through the use of a polarising beam splitter acting as a CNOT gate \cite{bib:PRL_80_3891}. The stochastic nature of any operations is also accounted by the loss metric, $p$.  Additionally, in any swapping procedure we assume that the inner two qubits are ideal, thus any error is transmitted through the channel and is only created by the error on the outer two qubits.

We can also move on to describe the the swapping operation, $S$, defined as \cite{bib:Pan01}
\begin{align}
        S(f_1,f_2) &= [f_1f_2+(1-f_1)(1-f_2)].
\end{align}
Which has the algebraic properties \cite{PhysRevA.57.822}
\begin{align}
    S(f_1,S(f_2,f_3)) &= S(S(f_1,f_2),f_3),\nonumber \\
    S(f_1,1) &= f_1,\nonumber \\
    S(f_1,f_2) &= S(f_2,f_1),\nonumber\\
    S\left(f_1,\frac{2f_1-4}{2f_1-1}\right) &= 1
\end{align}
Where $f_1,f_2,f_3$ are fidelities of individual channels in the network.  Therefore, swapping forms an Abelian group in the domain $[0,1]\textbackslash\frac{1}{2}$, suggesting that there exists a unique reduction for any set of linear edges regardless of order.  Furthermore, we can infer that performing swapping on two states, each with a fidelity less than 1, results in an overall fidelity less than the either of the original two fidelities, suggesting that the number of swapping operation performed should be minimized \cite{bib:NielsenChuang00}. 
\subsection{Entanglement Purification}
The introduction of entanglement purification establishes a unique feature of quantum networks that is not present in classical networks. The purpose of purification is that two low fidelity Bell pairs can be reduced into a single higher fidelity one \cite{bib:Pan01,bib:dur07}.  This can be done recursively allowing bell pairs to be raised back to an arbitrarily high fidelity. However, as this does require two Bell states the process scales at $2^n$ where $n$ is the number of bell pairs needed.  This operation  can be performed  using polarizing beam splitters, where it is stochastic.

At this point the unique phenomena of quantum networks begins to emerge.  While in classical network the overall idea is to minimize the length of a path through the graph, since this conveys the least amount of loss.  However, while this holds true for quantum networks, we are also attempting to maximize width since this will allow us to maximise the fidelity of our state. 

To prove it I define the effect of purification on fidelity as \cite{bib:Pan01}
\begin{align} \label{eq:F_P_purification}
    P(f_1,f_2) &= \frac{f_1f_2}{f_1f_2+(1-f_1)(1-f_2)},  
\end{align}
which possesses the algebraic properties
\begin{align}
P(f_1,f_2) &= P(f_2,f_1), \nonumber \\
P(P(f_1,f_2),f_3) &= P(f_1,P(f_2,f_3)), \nonumber \\
P(f_1,1/2) &= f_1, \nonumber\\
P(f_1,1-f_1) &= \frac{1}{2}
\end{align}
Which forms an abelian group under $[0,1]$, allowing purification across all qubits at a node to have a unique fidelity. At this point , for sake of completion, the algebraic properties of the logarithmic loss is
\begin{align}
\varepsilon(p_1,p_2) &= \varepsilon(p_2,p_1), \nonumber \\
\varepsilon(\varepsilon(p_1,p_2),p_3) &= \varepsilon(p_1,\varepsilon(p_2,p_3)),\nonumber\\
\varepsilon(p_1,1) &= p_1,
\end{align}
which is an abelian monoid. As with classical loss, this provides a limit on the effective distance of transfer.
\section{Area Laws}
\label{Area Laws}
\begin{figure}
\centering
\includegraphics[scale=0.2]{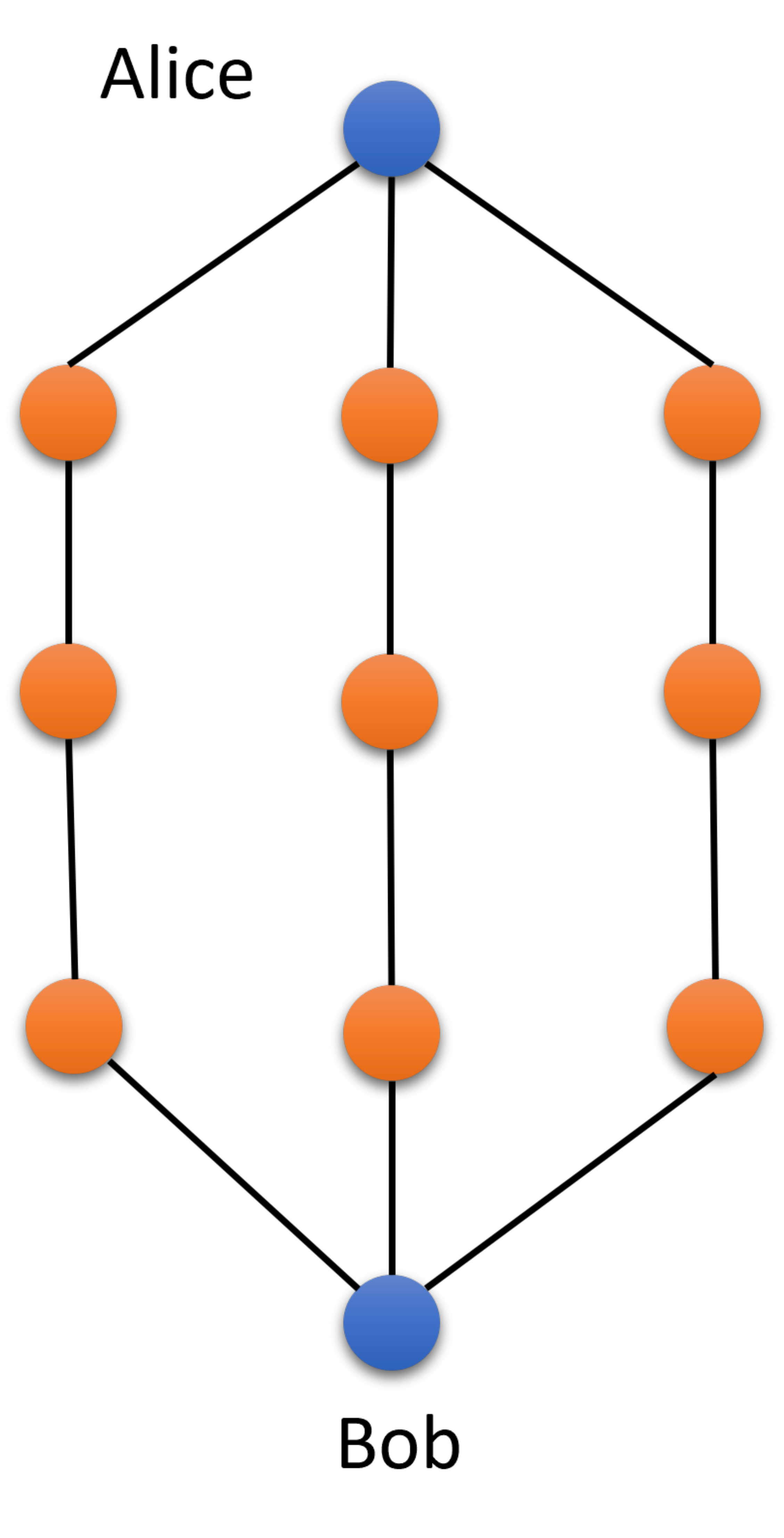}
\caption[A graph demonstrating a simple graph with a unique reduction.]{A graph demonstrating a simple graph with a unique reduction. Here the two users, Alice and Bob have some number of non-interacting paths between them, through which they can communicate.}
\end{figure}
At this point I have provided the formalism  to properly analyze various properties and scaling for a dephasing network. To begin, consider a multi-edge graph with two users Alice and Bob, with some number of independent paths between them consisting of a variety of nodes, as seen in Figure \ref{fig:simple graph}.  In the simplest case, all bell pairs go directly from Alice to Bob, thus the only operation that is performed is purification. The equation for fidelity of n bell pairs being purified is
\begin{align}
F = \frac{\prod_{i=1}^n F_i}{\prod_{i=1}^n F_i+\prod_{i=1}^n (1-F_i)}.
\end{align}
Expanding to consider $D$ nodes capable of performing swapping between Alice and Bob and $b$ pair capable of being purified at the endpoint, the log-Fidelity is
\begin{align}
    \log(F) &= \sum^\infty_{n=0} (-1)^n \log\left(\prod_{i=1}^b F_i-\prod_{i=1}^b (1-F_i)\right)^{D-n}\nonumber\\
    &\cdot \log\left(\prod_{i=1}^b F_i+\prod_{i=1}^b (1-F_i)\right)^n
\end{align}
Where the log of fidelity was chosen in order to provide a clearer picture of scaling.  As demonstrated the fidelity scales on the order of $\mathcal{O}(b^D)$, suggesting that that the number of purification operation should be maximized while the number of swapping is minimized. The scaling for the probability of success $P$ is $P_{\text{tot}}=P^{bd}$.  Therefore, any increase in operation decreases the probability of success.  Thus, any optimization problem needs to find the balance between the success probability and increasing the number of purifications performed.  In addition, the number of swapping operations should be kept to a minimum in order to maximize fidelity and chance of success.
\subsection{Limits of Reduction}
\begin{figure}
\centering
\includegraphics[scale=0.2]{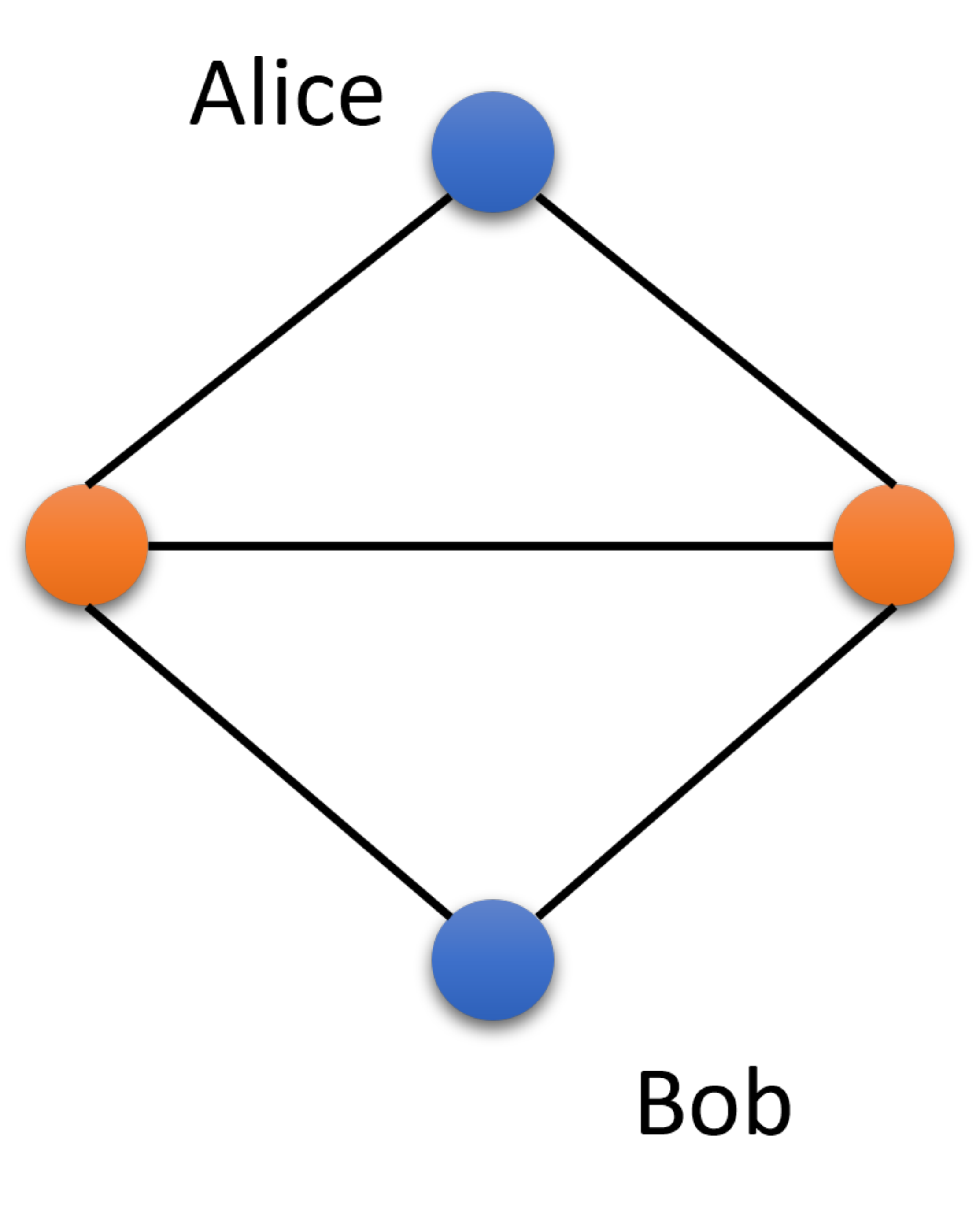}
\caption[A graph with a non-unique reduction.]{A graph with a non-unique reduction. Note that purification can be performed in two places, one is at one of the center nodes where one path is create by Alice directly to that node, with the other going through the other middle node and through the center connection.  The other option is to ignore the center connection all together reducing the graph to the simplest case of the graph shown in Figure \ref{fig:simple graph}}
\label{fig:Wheatstone}
\end{figure}
It is important to note that these rules do not hold for all cases.  The main point being that the algebra only holds for simple dephasing.  In other decoherence models the abelian nature of the swapping and purification operations breaks down due to the methods used to compensate.  For example, Pauli channels require the addition of twirling which prevents the associative feature from occurring \cite{bib:dur98}.

Furthermore, a unique reduction only exists for simple cases where each path between the two points is independent. For example, consider the graph in Figure \ref{fig:Wheatstone}, here there exists multiple choices for how the graph can be reduced.  For this specific case it must be decided whether the middle edge should be ignored, reducing the transmission to Figure \ref{fig:simple graph}, or whether purification should be performed at the left or right node before being sent to the endpoint.  Therefore, the ideal reduction must simply be solved for by trying every combination. Fortunately, the abelian nature of the swapping and purification operations the actual problem of reducing a given network to a single fidelity is isomorphic to simplifying a resistor network using the series and parallel resistor reduction provided by Ohm's law. However, do to the lack of a current in cases such as Figure \ref{fig:Wheatstone} lack a unique answer.  
\subsection{Optimal Subgraphs}
With the rules and restrictions provide enough restrictions to create a rudimentary algorithm capable of finding the optimal subgraph for a transmission on the network.  To begin, assume that there is only one pair communicating on the network.  While, the rules outlined and algorithm proceeding will work for a multi-user circumstance, additional users require optimal partitioning to be done, requiring additional costs to perform optimal user partitions. Approaches to the multi-user can be found in the original paper at \cite{leone2021qunet}.  However, for a single case the partitioning problem can be performed simply by deciding on an optimal success rate.  At this point, select purely swapping paths that are above the threshold value, this can be done by performing multiple iterations of Dijkstra's algorithm \cite{bib:Dijkstra59}.  The resulting subgraph will provide the space to path through. The graph can then be simplified by repeated applications of the swapping and purification operations, similar to the reduction of a circuit network and thus will be brought to an irreducible state in the order of $\mathcal{O}(n\text{log}(n))$ \cite{DUFFIN1965303}.  The remaining network is either the ideal transmission if it is fully reduced, or lacks a unique answer and must be searched for the optimal answer.
\section{Conclusion}
In conclusion, the routing problem differs significantly between classical and quantum networks.  The first is that in a classical network we aim to find the shortest path, whereas in a quantum network the main issue is to find the optimal subgraph. As such we seek to maximize breadth of a given subgraph while minimizing depth, since quantum networks maintain fidelity by distributing information through redundant routes and purifying to be able to obtain the final results. In terms of fidelity, adding resources in the form of additional paths to purify with cannot reduce performance. Though this must be balanced with the chance of success.   Ultimately, this work lays down basic design principles and considerations to be accounted for when designing quantum networks \cite{rohde2021quantum,khatri_spooky_2021,PhysRevA.97.012335}.

\chapter*{Chapter 4. \\ Quantum Imaging}
\addcontentsline{toc}{chapter}{Chapter 4. Quantum Imaging}
\setcounter{section}{0}
\setcounter{figure}{0}
\setcounter{table}{0}
\refstepcounter{chapter}
\label{QI}
\doublespacing
The spatial resolution of optical imaging systems is established by the diffraction of photons and the noise associated with their quantum fluctuations \cite{abbe1873beitrage, rayleigh1879xxxi, born2013principles, goodman2005introduction, magana2019quantum}. For over a century, the Abbe-Rayleigh criterion has been used to assess the diffraction-limited resolution of optical instruments \cite{born2013principles, won2009eyes}. At a more fundamental level, the ultimate resolution of optical instruments is established by the laws of quantum physics through the Heisenberg uncertainty principle \cite{stelzer2002beyond, kolobov2000quantum, stelzer2000uncertainty}. In classical optics, the Abbe-Rayleigh resolution criterion stipulates that an imaging system cannot resolve spatial features smaller than $\lambda/2\text{NA}$. In this case, $\lambda$ represents the wavelength of the illumination field, and $\text{NA}$ describes numerical aperture of the optical instrument \cite{abbe1873beitrage, rayleigh1879xxxi, born2013principles, editorial2009}.  Given the implications that overcoming the Abbe-Rayleigh resolution limit has for multiple applications, such as, microscopy, remote sensing, and astronomy \cite{pirandola2018advances, hell20152015, editorial2009, born2013principles}, there has been an enormous interest in improving the spatial resolution of optical systems \cite{tsang2009quantum, tsang2016quantum, hell1994breaking}. So far, optical superresolution has been achieved through spatial decomposition of eigenmodes \cite{paur2018tempering, tsang2016quantum, tamburini2006overcoming}. These conventional schemes rely on spatial projective measurements to pick up phase information that is used to boost spatial resolution of optical instruments \cite{tsang2016quantum,Steinberg17PRL,zhou2019quantum,Treps20Optica,Saleh18Optica,Liang21Optica}.

For almost a century, the importance of phase over amplitude information has constituted established knowledge for optical engineers \cite{goodman2005introduction, born2013principles,magana2019quantum}. Recently, this idea has been extensively investigated in the context of quantum metrology \cite{boto2000quantum, tang2016fault, parniak2018beating,you2021scalable, magana2019quantum}.  More specifically, it has been demonstrated that phase information can be used to surpass the Abbe-Rayleigh resolution limit for the spatial identification of light sources \cite{tsang2009quantum, giovannetti2009sub, Steinberg17PRL, zhou2019quantum, Treps20Optica}. For example, phase information can be obtained through mode decomposition by using projective measurements or demultiplexing of spatial modes \cite{tsang2016quantum, tamburini2006overcoming, Steinberg17PRL, zhou2019quantum, Treps20Optica}. Naturally, these approaches require \textit{a priori} information regarding the coherence properties of the, in principle, “unknown” light sources \cite{hell1994breaking, Saleh18Optica, Liang21Optica, tsang2016quantum}. Furthermore, these techniques impose stringent requirements on the alignment and centering conditions of imaging systems \cite{tamburini2006overcoming, Steinberg17PRL, zhou2019quantum, Treps20Optica, hell1994breaking, Saleh18Optica, OmarSciAdv2016, YangLight2017, Liang21Optica, tsang2016quantum}. Despite these limitations, most, if not all, the current experimental protocols have relied on spatial projections and demultiplexing in the Hermite-Gaussian, Laguerre-Gaussian, and parity basis \cite{tamburini2006overcoming, zhou2019quantum, Steinberg17PRL, Saleh18Optica, zhou2019quantum, Treps20Optica, Liang21Optica, tsang2016quantum}.

\begin{figure*}[!t]
  \centering
 \includegraphics[width=1\textwidth]{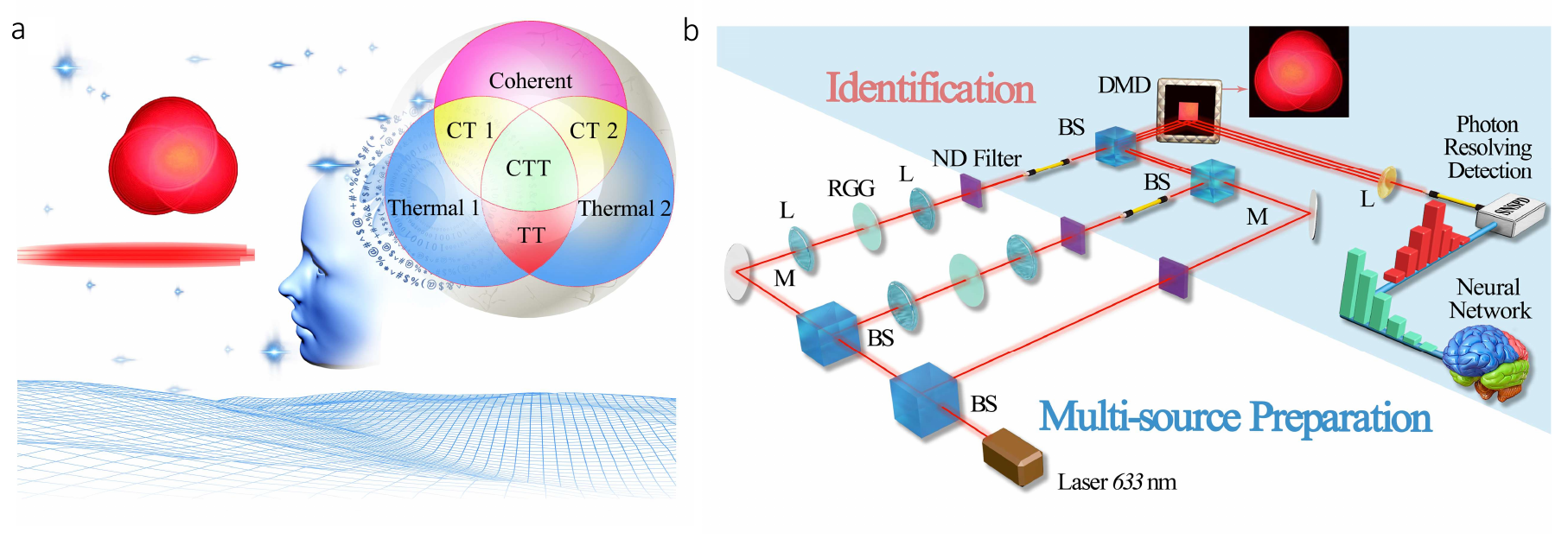}
\caption[Conceptual illustration and schematic of our experimental setup to demonstrate superresolving imaging.]{Conceptual illustration and schematic of our experimental setup to demonstrate superresolving imaging. 
The illustration in \textbf{a} depicts a scenario where diffraction limits the resolution of an optical instrument for remote imaging. In our protocol, an artificial neural network enables the identification of the photon statistics that characterize the point sources that constitute a target object. In this case, the point sources emit either coherent or thermal photons. Remarkably, the neural network is capable of identifying the corresponding photon fluctuations and their combinations, for example coherent-thermal (CT1, CT2), thermal-thermal (TT) and coherent-thermal-thermal (CTT). This capability allows us to boost the spatial resolution of optical instruments beyond the Abbe-Rayleigh resolution limit.
The experimental setup in \textbf{b} is designed to generate two independent thermal and one coherent light sources.  The three sources are produced from a continuous-wave (CW) laser at $633 \text{ nm}$. The CW laser beam is divided by two beam splitters (BS) to generate three spatial modes, two of which are then passed through rotating ground glass (RGG) disks to produce two independent thermal light beams. The three light sources, with different photon statistics, are attenuated using neutral density (ND) filters and then combined to mimic a remote object such as the one shown in the inset of \textbf{b}. This setup enables us to generate multiple sources with tunable statistical properties.  The generated target beam is then imaged onto a digital micro-mirror device (DMD) that we use to perform raster scanning. The photons reflected off the DMD are collected and measured by a single-photon detector. Our protocol is formalized by performing photon-number-resolving detection \cite{you2020identification}. The characteristic quantum fluctuations of each light source are identified by an artificial neural network. This information is then used to produce a high-resolution image of the object beyond the diffraction limit. }
\label{schematic}
\end{figure*}

The quantum statistical fluctuations of photons  establish the nature of light sources \cite{You2021naturecomm,mandel78,magana2019multiphoton,you2020identification, gerry2005introductory}. As such, these fundamental properties are not affected by the spatial resolution of an optical instrument \cite{gerry2005introductory}. Here,  I demonstrate that measurements of the quantum statistical properties of a light field enable imaging beyond the Abbe-Rayleigh resolution limit. This is performed by exploiting the self-learning features of artificial intelligence to identify the statistical fluctuations of photon mixtures \cite{you2020identification}. More specifically, I demonstrate a smart quantum camera with the capability to identify photon statistics at each pixel. For this purpose, I introduce a universal quantum model that describes the photon statistics produced by the scattering of an arbitrary number of light sources. This model is used to design and train artificial neural networks for the identification of light sources. Remarkably, this scheme overcomes inherent limitations of existing superresolution protocols based on spatial mode projections and multiplexing \cite{tsang2016quantum, tamburini2006overcoming, zhou2019quantum, Steinberg17PRL, Saleh18Optica,  Treps20Optica, Liang21Optica}.  This work was originally publiched in Bhusal et al \cite{bhusal2021smart}.

\section{Theory}
The conceptual schematic behind the experiment is depicted in 
Fig. \ref{schematic}\textbf{a}. This camera utilizes an artificial neural network to identify the photon statistics of each point source that constitutes a target object. The description of the photon statistics produced by the scattering of an arbitrary number of light sources is achieved through a general model that relies on the quantum theory of optical coherence introduced by Sudarshan and Glauber \cite{sudarshan1963equivalence, glauber1963quantum, gerry2005introductory}. This model is applied to
design and train a neural network capable of identifying light sources at each pixel of our camera. This unique feature is achieved by performing photon-number-resolving detection \cite{you2020identification}. The sensitivity of this camera is limited by the photon fluctuations, as stipulated by the Heisenberg uncertainty principle, and not by the Abbe-Rayleigh resolution limit \cite{gerry2005introductory, magana2019quantum}. 

\begin{figure*}[!ht]
\centering
\includegraphics[width=1\textwidth]{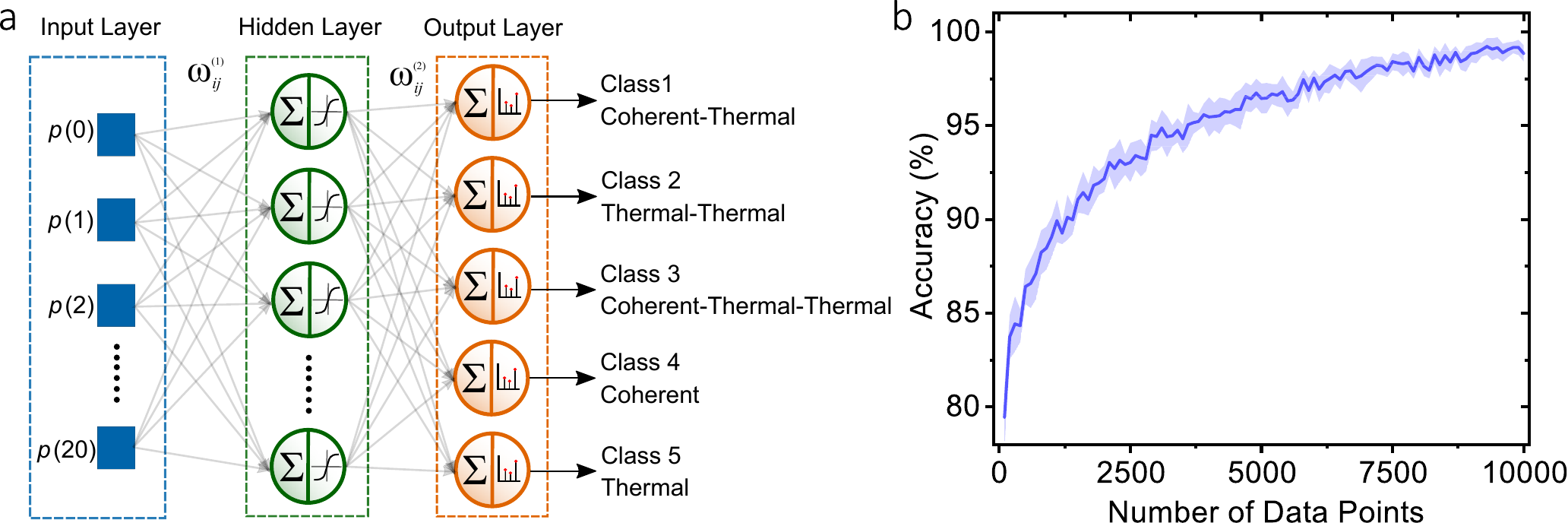}
\caption[The scheme of the two-layer neural network used to identify the photon statistics produced by a combination of three sources and it's accuracy.]{The scheme of the two-layer neural network used to identify the photon statistics produced by a combination of three sources is shown in \textbf{a}. The computational model consists of an input layer, a hidden layer of sigmoid neurons, and a Softmax output layer. The training of the neural network through Eqs. \ref{pthcoh} and \ref{pthcohdis} enables the efficient identification of five classes of photon statistics. Each class is characterized by a $g^{(2)}$ function, which is defined by a specific combination of light sources \cite{you2020identification}. In this experiment, these classes correspond to the characteristic photon statistics produced by coherent or thermal light sources and their combinations. For example, coherent-thermal, thermal-thermal, or coherent-thermal-thermal.  The figure in \textbf{b} shows the performance of the neural network as a function of the number of data samples used each time in the testing process. The classification accuracy for the five possible complex classes of light is 80\% with 100 data points. Remarkably, the performance of the neural network increases to approximately 95\% when we use 3500 data points in each test sample.}
\label{nnAccuracy}
\end{figure*}
In general, realistic imaging instruments deal with the detection of multiple light sources. These sources can be either distinguishable or indistinguishable \cite{born2013principles, gerry2005introductory}. The combination of indistinguishable sources can be represented by either coherent or incoherent superpositions of light sources characterized by Poissonian (coherent) or super-Poissonionan (thermal) statistics \cite{gerry2005introductory}. In my model, I first consider the indistinguishable detection of $N$ coherent and $M$ thermal sources. For this purpose, I make use of the P-function  $P_{\text{coh}}(\gamma)=\delta^2 (\gamma-\alpha_{k})$ to model the contributions from the $k$th coherent source with the corresponding complex amplitude $\alpha_k$ \cite{sudarshan1963equivalence, glauber1963quantum}. The total complex amplitude associated to the superposition of an arbitrary number of light sources is given by $\alpha_{\text{tot}}=\sum_{k=1}^{N}\alpha_k$.
In addition, the P-function for the $l$th thermal source, with the corresponding mean photon numbers $\bar{m}_l$, is defined as $P_{\text{th}}(\gamma)=(\pi\bar{m}_l)^{-1}\exp{(-|\gamma|^2}/\bar{m}_l)$. The total number of photons attributed to the $M$ number of thermal sources is defined as $m_{\text{tot}}=\sum_{l=1}^{M}\bar{m}_l$. 
These quantities allow me to calculate the P-function for the multisource system as

\begin{equation}
\label{pfun}
\begin{aligned}
P_{\text{th-coh}}(\gamma)& = \int\cdots\int  P_{N+M}(\gamma-\gamma_{N+M-1})\\
&  \times \cor{\prod_{i=2}^{N+M-1} P_i(\gamma_i-\gamma_{i-1})d^{2}\gamma_i} P_{1}(\gamma_{1})d^{2}\gamma_{1}.
\end{aligned}
\end{equation}
This approach enables the analytical description of the photon-number distribution $p_{\text{th-coh}}(n)$ associated to the detection of an arbitrary number of indistinguishable light sources. This is calculated as $p_{\text{th-coh}}(n)=\bra{n} \hat{\rho}_{\text{th-coh}} \ket{n}$, where $\rho_{\text{th-coh}}=\int{P_{\text{th-coh}}(\gamma) \ket{\gamma}\bra{\gamma}}d^2\gamma$. After algebraic manipulation (see Appendix \ref{app1}), I obtain the following photon-number distribution
\begin{equation}
\label{pthcoh}
\begin{aligned}
p_{\text{th-coh}}(n)=&\frac{\left(m_{\text{tot}}\right)^{n} \exp \left(-\left(|\alpha_\text{tot}|\right)^{2} / m_{\text {tot}}\right)}{\pi\left(m_{\text{tot}}+1\right)^{n+1}}\\
& \times \sum_{k=0}^{n} \frac{1}{k !(n-k) !} \Gamma\left(\frac{1}{2}+n-k\right) \Gamma\left(\frac{1}{2}+k\right)\\
& { }_{1} F_{1}\left(\frac{1}{2}+n-k; \frac{1}{2}; \frac{(\operatorname{Re}[\alpha_\text{tot}])^{2}}{m_{\text{tot}}\left(m_{\text{tot}}+1\right)}\right)\\
& { }_{1} F_{1}\left(\frac{1}{2}+k;\frac{1}{2}; \frac{(\operatorname{Im}[\alpha_\text{tot}])^{2}}{m_{\text{tot}}\left(m_{\text{tot}}+1\right)}\right),
\end{aligned}
\end{equation}
\\
\\
where $\Gamma(z)$ and ${ }_{1} F_{1}(a;b;z)$ are the Euler gamma and the Kummer confluent hypergeometric functions, respectively. This probability function enables the general description of the photon statistics produced by any indistinguishable combination of light sources. Thus, the photon distribution produced by the distinguishable detection of $N$ light sources can be simply obtained by performing a discrete convolution of Eqn. \ref{pthcoh} as
\begin{equation}
\label{pthcohdis}
\begin{aligned}
p_{\text{tot}}(n)=&\sum_{{m_1}=0}^{n} \sum_{{m_2}=0}^{n-m_1}  \dotsb \sum_{{m_{N-1}}=0}^{n-\sum^{N-1}_{j=1}m_j} p_1(m_1)p_2(m_2)\dotsb \\
& p_{N-1}(m_{N-1}) p_N(n-\sum^{N-1}_{j=1}m_j).
\end{aligned}
\end{equation}
The combination of Eqn. \ref{pthcoh} and Eqn. \ref{pthcohdis} allows the classification of photon-number distributions for any combination of light sources. 
\section{Experiment}
The proof-of-principle quantum camera is demonstrated using the experimental setup shown in Fig. \ref{schematic}\textbf{b}. For this purpose,  a continuous-wave laser at $ 633 \text{nm}$ is used to produce either coherent, or incoherent superpositions of distinguishable, indistinguishable, or partially distinguishable light sources.  In this case, the combination of photon sources, with tunable statistical fluctuations, acts as the target object. Then, the target object is imaged onto a digital micro-mirror device (DMD) that is used to implement raster scanning. This is implemented by selectively turning on and off groups of pixels in the DMD. The light reflected off the DMD is measured by a single-photon detector that performs photon-number-resolving detection. This is implemented through the technique described in ref. \cite{you2020identification}.

\begin{figure}[ht!]
\centering
\includegraphics[scale=.7]{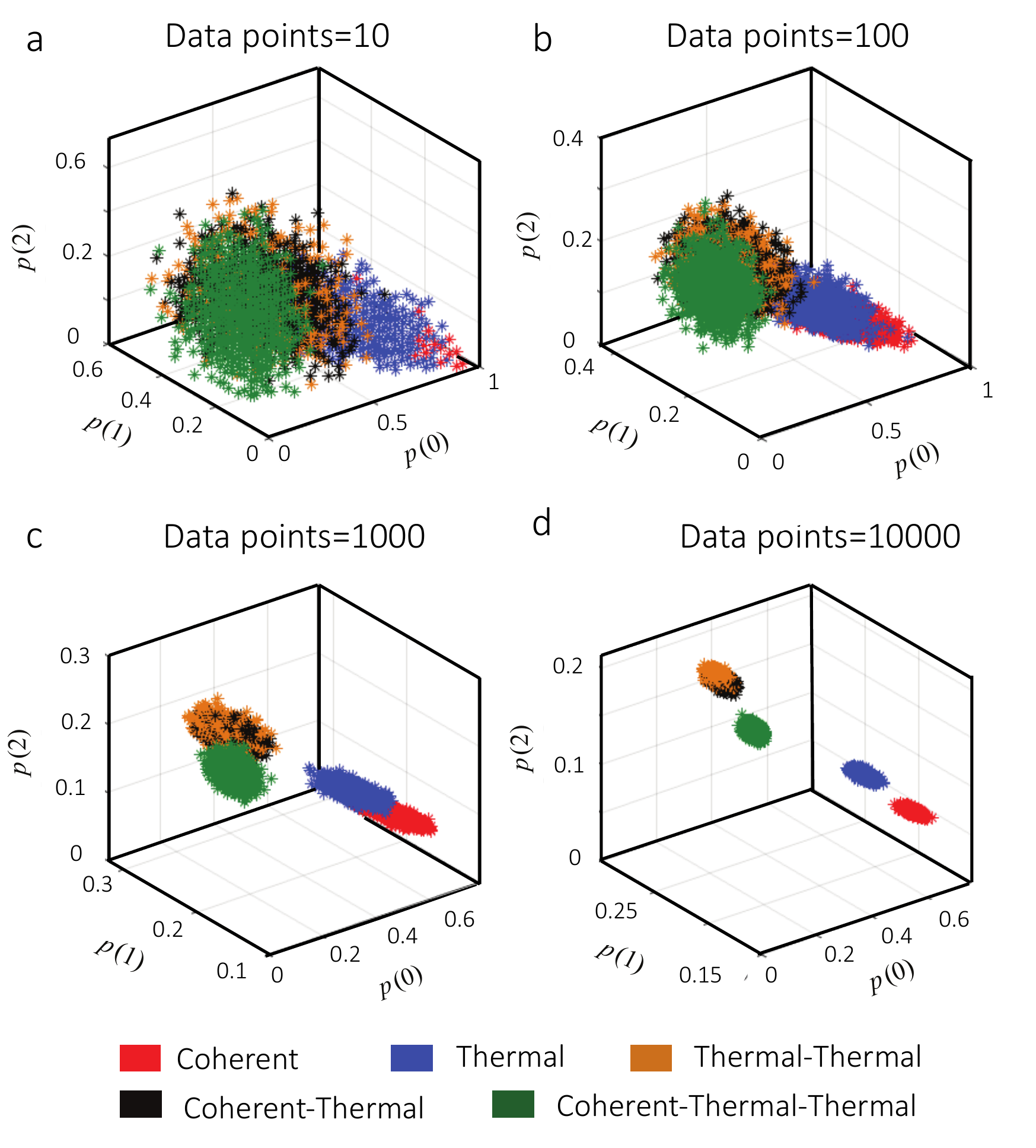}
\caption[Projection of the feature space on the plane defined by the probabilities $p(0)$, $p(1)$, and $p(2)$.]{Projection of the feature space on the plane defined by the probabilities $p(0)$, $p(1)$, and $p(2)$. The red points correspond to the photon statistics for coherent light, and the blue points indicate the photon statistics for thermal light fields. Furthermore, the brown dots represent the photon statistics produced by the scattering of two thermal light sources, and the black points show the photon statistics for a mixture of photons emitted by one coherent and one thermal source. The corresponding statistics for a mixture of one coherent and two thermal sources are indicated in green. As shown in \textbf{a}, the distributions associated to the multiple sources obtained for 10 data points are confined to a small region of the feature space. A similar situation prevails in \textbf{b} for 100 data points. As shown in panel \textbf{c}, the distributions produced with 1000 data points occupy different regions, although brown and black points keep closely intertwined. Finally, the separated distributions obtained with 10000 data points in \textbf{d} enable efficient identification of light sources.}
\label{fig:numberdatapoint}
\end{figure}

\begin{figure*}[!htbp]
  \centering
 \includegraphics[width=0.95\textwidth]{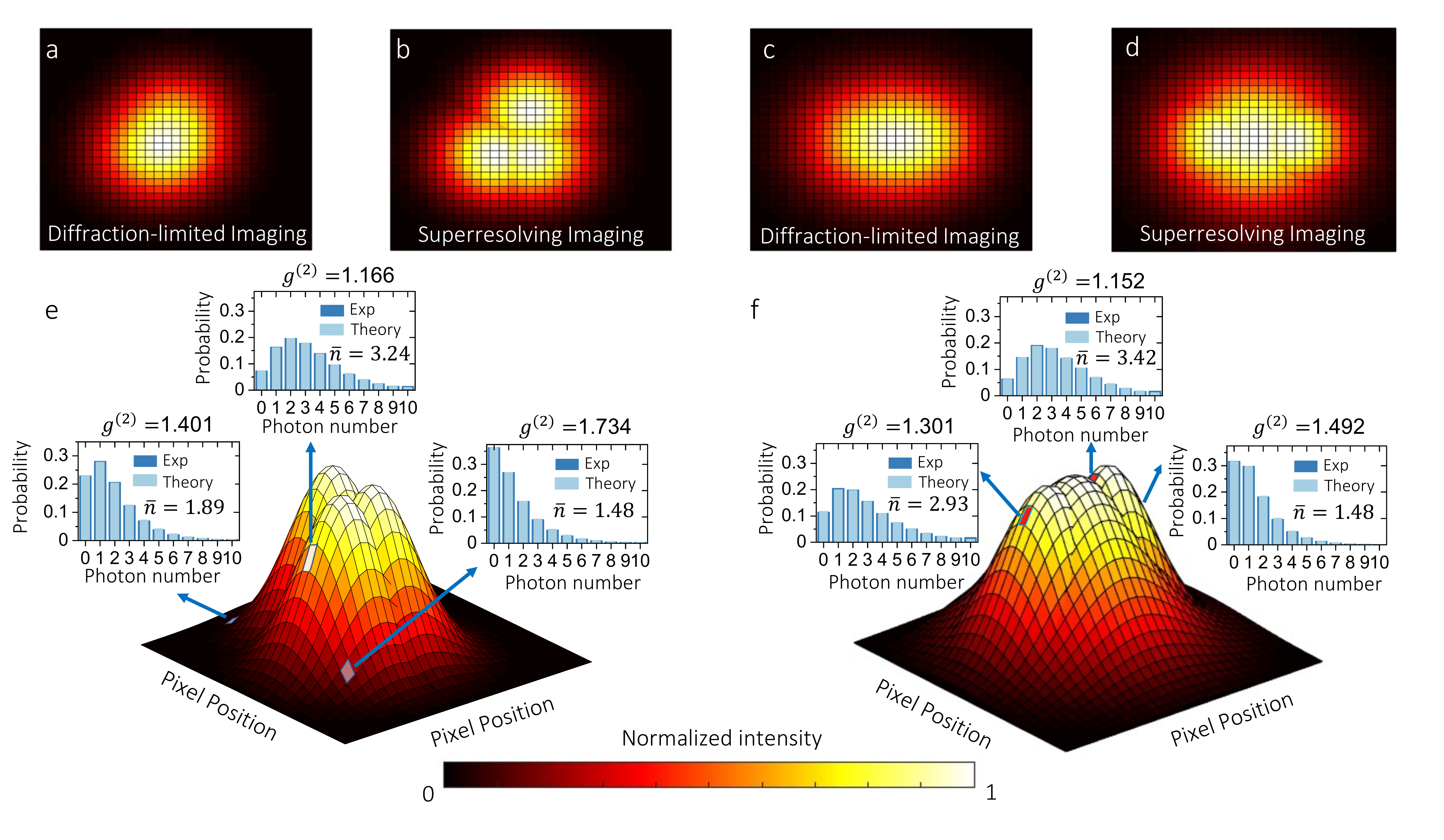}
\caption[Experimental superresolving imaging.]{Experimental superresolving imaging. The plot in \textbf{a} shows the combined intensity profile of the three partially distinguishable sources. As stipulated by the Abbe-Rayleigh resolution criterion, the transverse separations among the sources forbid their identification. As shown in \textbf{b}, the smart quantum camera enables superresolving imaging of the remote sources.  In \textbf{c} and \textbf{d},  another experimental realization of the protocol for a different distribution of light sources is demonstrated. In this case, two small sources are located inside the point-spread function of a third light source. The figures in \textbf{e} and \textbf{f} correspond to the inferred spatial distributions based on the experimental pixel-by-pixel imaging used to produce \textbf{b} and \textbf{d}. The insets in \textbf{e} and \textbf{f} show photon-number probability distributions for three pixels, the theory bars were obtained through Eqs. \ref{pthcoh} and \ref{pthcohdis}. These results demonstrate the potential of the technique to outperform conventional diffraction-limited imaging.}
\label{profiles}
\end{figure*}

The equations above informs the implementation of a multi-layer feed-forward network for the identification of the quantum photon fluctuations of the point sources of a target object. The structure of the network consists of a group of interconnected neurons arranged in layers. Here, the information flows only in one direction, from input to output \cite{svozil1997introduction, bhusal2021spatial}. As indicated in Fig. \ref{nnAccuracy}\textbf{a}, the network comprises two layers, with ten sigmoid neurons in the hidden layer (green neurons) and five softmax neurons in the output layer (orange neurons). In this case, the input features represent the probabilities of detecting $n$ photons at a specific pixel, $p(n)$, whereas the neurons in the last layer correspond to the classes to be identified. The input vector is then defined by twenty-one features corresponding to $n$=0,1,...,20. In the experiment, I define five classes that I label as: coherent-thermal (CT), thermal-thermal (TT), coherent-thermal-thermal (CTT), coherent (C), and thermal (T). If the brightness of the experiment remains constant, these classes can be directly defined through the photon-number distribution described by Eqs. \ref{pthcoh} and \ref{pthcohdis}. However, if the brightness of the sources is modified, the classes can be defined through the $g^{(2)}=1+\left(\left\langle(\Delta \hat{n})^{2}\right\rangle-\langle\hat{n}\rangle\right) /\langle\hat{n}\rangle^{2}$, which is intensity-independent \cite{you2020identification, You2021naturecomm}. The parameters in the $g^{(2)}$ function can also be calculated from Eqs. \ref{pthcoh} and \ref{pthcohdis}. It is important to mention that the output neurons provide a probability distribution over the predicted classes \cite{Goodfellow2016,bishop2006pattern}. The training details of the neural networks can be found in Appendix \ref{app1}. 


I test the performance of our neural network through the classification of a complex mixture of photons produced by the combination of one coherent with two thermal light sources. The accuracy of the trained neural network is reported in Fig. \ref{nnAccuracy}\textbf{b}. In the setup, the three partially overlapping sources form five classes of light with different mean photon numbers and photon statistics. I exploit the functionality of our artificial neural network to identify the underlying quantum fluctuations that characterize each kind of light. I calculate the accuracy as the ratio of true positive and true negative to the total of input samples during the testing phase. Fig. \ref{nnAccuracy}\textbf{b} shows the overall accuracy as a function of the number of data points used to build the probability distributions for the identification of the multiple light sources using a supervised neural network. The classification accuracy for the mixture of three light sources is 80\% with 100 photon-number-resolving measurements. The performance of the neural networks increases to approximately 95\% when I use 3500 data points to generate probability distributions.

The performance of the protocol for light identification can be understood through the distribution of light sources in the probability space shown in Fig. \ref{fig:numberdatapoint}. Here, the projection of the feature space on the plane defined by the probabilities $p(0)$, $p(1)$, and $p(2)$ for different number of data points. Each point is obtained from an experimental probability distribution. As illustrated in Fig. \ref{fig:numberdatapoint}\textbf{a}, the distributions associated to the multiple sources obtained for 10 data points are confined to a small region of the feature space. This condition makes extremely hard the identification of light sources with 10 sets of measurements. A similar situation can be observed for the distribution in Fig. \ref{fig:numberdatapoint}\textbf{b} that was generated using 100 data points. As shown in panel Fig. \ref{fig:numberdatapoint}\textbf{c}, the separations in the distributions produced with 1000 data points occupy different regions, although brown and black points keep closely intertwined. These conditions enable one to identify multiple light sources. Finally, the separated distributions obtained with 10000 data points in Fig. \ref{fig:numberdatapoint}\textbf{d} enable efficient identification of light sources. These probability space diagrams explain the performances reported in Fig. \ref{nnAccuracy}. An interesting feature of Fig. \ref{fig:numberdatapoint} is the fact that the distributions in the probability space are linearly separable. 

\begin{figure*}[t!]
 \centering
 \includegraphics[width=0.95\textwidth]{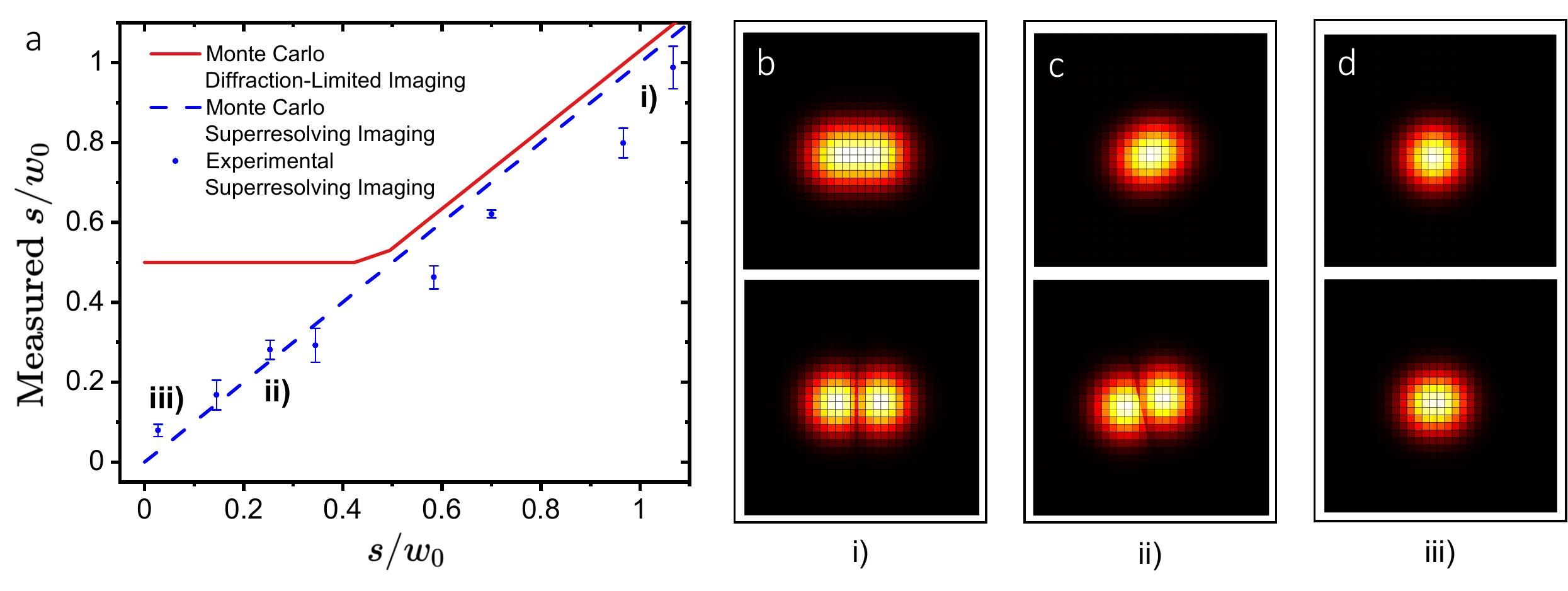}
 \caption[Comparison between the spatial resolution of the camera and direct imaging.]{Comparison between the spatial resolution of the camera and direct imaging. Here the distance is normalized by the beam radius for easy identification of the Abbe-Rayleigh limit. As shown in \textbf{a}, the red line is the result of a Monte-Carlo simulation for traditional intensity based direct imaging.  The plateau is the area where the algorithm becomes unstable. The dotted blue line represents the limit for our supperresolving imaging method, where perfect classification of each pixel is assumed. The blue dots represent the experimental data collected with the camera for superresolving imaging. The experimental points demonstrate the potential of the technique for identifying spatial features beyond the Abbe-Rayleigh resolution criterion. The 
 first row in the panels from \textbf{b} to \textbf{d} shows the reconstructed spatial profiles 
 obtained through direct imaging whereas the second row shows the superresolving images obtained with our technique. The panel in \textbf{b} shows the spatial profiles for the experimental point $i)$. This corresponds to the experimental detection of two sources with the largest separation. The spatial profiles in \textbf{c} correspond to the experimental point labeled as $ii)$. Finally, the panel in \textbf{d} shows the spatial distributions for the experimental point with the smallest separation, this is labeled as $iii)$.}
 \label{distances}
\end{figure*}

As demonstrated in Fig. \ref{profiles}, the identification of the quantum photon fluctuations at each pixel of the camera enables me to demonstrate superresolving imaging.  In the experiment each source has a mean photon number between 1 and 1.5 for the brightest pixel. The raster-scan image of a target object composed of multiple partially distinguishable sources in Fig. \ref{profiles}\textbf{a} illustrates the performance of conventional imaging protocols limited by diffraction \cite{goodman2005introduction, won2009eyes, stelzer2002beyond, kolobov2000quantum}. In this case, it is practically impossible to identify the multiple sources that constitute the target object. Remarkably, as shown in Fig. \ref{profiles}\textbf{b}, the protocol provides a dramatic improvement of the spatial resolution of the imaging system. In this case, it becomes clear the presence of the three emitters that form the remote object. The estimation of separations among light sources is estimated through a fit over the classified pixel-by-pixel image. Additional details can be found in Appendix \ref{app1}. In Figs. \ref{profiles}\textbf{c} and \textbf{d}, I demonstrate the robustness of our protocol by performing superresolving imaging for a different configuration of light sources. In this case, two small sources are located inside the point-spread function of a third light source. As shown in Fig. \ref{profiles}\textbf{c}, the Abbe-Rayleigh limit forbids the identification of light sources. However, I demonstrate substantial improvement of spatial resolution in Fig. \ref{profiles}\textbf{d}. The plots in Figs. \ref{profiles}\textbf{e} and \textbf{f} correspond to the inferred spatial distributions based on the experimental pixel-by-pixel imaging used to produce Figs. \ref{profiles}\textbf{b} and \textbf{d}. The insets in Figs. \ref{profiles}\textbf{e} and \textbf{f} show photon-number probability distributions for three pixels. The theoretical photon-number distributions in Fig.\ref{profiles}\textbf{e} and \textbf{f} are obtained through a procedure of least square regression \cite{massaron2016regression}. Here the least squares difference between the measured and theoretical probability distribution was minimized for $0\leq n\leq 6$. The sources were assumed to be partially distinguishable  allowing the theoretical distribution to be defined by Eqs. \ref{pthcoh} and Eqn. \ref{pthcohdis}.  The combined mean photon numbers of each source generated for the fit totals the measured mean photon number (see Appendix \ref{app1}). This scheme enables the use of the photon-number distributions or their corresponding $g^{(2)}$ to characterize light sources. This allows me to determine each pixel's corresponding statistics, regardless of the mean photon numbers of the sources in the detected field \cite{you2020identification, You2021naturecomm}.
\section{Surpassing the Abbe-Rayleigh Criterion}
I now provide a quantitative characterization of the superresolving imaging scheme based on the identification of photon statistics. I demonstrate that our smart camera for superresolving imaging can capture small spatial features that surpass the resolution capabilities of conventional schemes for direct imaging \cite{abbe1873beitrage, rayleigh1879xxxi, born2013principles, goodman2005introduction, magana2019quantum}. Consequently, as shown in Fig. \ref{distances}, the camera enables the possibility of performing imaging beyond the Abbe-Rayleigh criterion. In this case,  multiple experiments were performed in which a superposition of partially distinguishable sources were imaged. The superposition was prepared using one coherent and one thermal light source. In Fig. \ref{distances}\textbf{a}, I plot the predicted transverse separation $s$ normalized by the Gaussian beam waist radius $w_0$ for both protocols. Here $w_0=\lambda/\pi\text{NA}$, this parameter is directly obtained from the experiment.   As demonstrated in Fig. \ref{distances}\textbf{a}, the protocol enables one to resolve spatial features for sources with small separations even for diffraction-limited conditions. As expected for larger separation distances, the performance of the protocol matches the accuracy of intensity measurements. This is further demonstrated by the spatial profiles shown from Fig. \ref{distances}\textbf{b} to \textbf{d}. The first row shows spatial profiles for three experimental points in Fig. \ref{distances}\textbf{a} obtained through direct imaging whereas the images in the second row were obtained using our scheme for superresolving imaging. The spatial profiles in Fig. \ref{distances}\textbf{b} show that both imaging techniques lead to comparable resolutions and the correct identification of the centroids of the two sources. However, as shown in Fig. \ref{distances}\textbf{c} and \textbf{d}, the camera outperforms direct imaging when the separations decrease. Here, the actual separation is smaller than $w_0/2$ for both cases. It is worth noticing that in this case, direct imaging cannot resolve spatial features of the sources. Here, the predictions of direct imaging become unstable and erratic. Remarkably, the simulations show an excellent agreement with the experimental data obtained for our scheme for superresolving imaging (see Appendix \ref{app1}).
\section{Conclusion}
In conclusion, I demonstrated a robust quantum camera that enables superresolving imaging beyond the Abbe-Rayleigh resolution limit. Our scheme for quantum statistical imaging exploits the self-learning features of artificial intelligence to identify the statistical fluctuations of truly unknown mixtures of light sources. This particular feature of the scheme relies on a universal model based on the theory of quantum coherence to describe the photon statistics produced by the scattering of an arbitrary number of light sources. I demonstrated that the measurement of the quantum statistical fluctuations of photons enables one to overcome inherent limitations of existing superresolution protocols based on spatial mode projections \cite{tsang2016quantum,Steinberg17PRL,zhou2019quantum,Treps20Optica,Saleh18Optica,Liang21Optica}. We believe that this work represents a new paradigm in the field of optical imaging with important implications for microscopy, remote sensing, and astronomy \cite{magana2019quantum, won2009eyes, stelzer2002beyond, kolobov2000quantum, stelzer2000uncertainty,editorial2009, pirandola2018advances}.

\singlespacing
\chapter*{Chapter 5. \\ Multiphoton Quantum van Cittert-Zernike Theorem}
\addcontentsline{toc}{chapter}{Chapter 5. Multiphoton Quantum van Cittert-Zernike Theorem}
\setcounter{section}{0}
\setcounter{figure}{0}
\setcounter{table}{0}
\refstepcounter{chapter}
\label{MQvCZT}
\doublespacing
The van Cittert-Zernike theorem constitutes one of the pillars of optical physics \cite{Cittert:34,ZERNIKE:38}. As such, this fundamental theorem provides the  formalism to describe the modification of the coherence properties of optical fields upon propagation \cite{Cittert:34,ZERNIKE:38,born2013principles,Wolf:54}. Over the last decades, extensive investigations have been conducted to explore the evolution of spatial, temporal, spectral, and polarization coherence of diverse families of optical beams \cite{Dorrer:04, gori2000use, cai2020, Cai2012}. In the context of classical optics, the extensive investigation of the van Cittert-Zernike theorem led to the development of schemes for optical sensing, metrology, and astronomical interferometry \cite{Carozzi:09, Batarseh:18, Barakat:2000}. Nowadays, there has been an enormous impetus to explore the implications of the van Cittert-Zernike theorem for quantum mechanical systems \cite{Saleh05PRL,Fabre_2017, Howard2019,Barrachina2020}. Recent efforts have been devoted to study the evolution of the properties of spatial coherence of biphoton systems \cite{Saleh05PRL, Howard2019, PhysRevA.95.063836, PhysRevA.99.053831}. Interestingly, this research unveiled similarities between coherence and entanglement \cite{Saleh05PRL,Qian:18,Eberly_2016}. Moreover, the possibility of describing the evolution of spatial coherence and entanglement of propagating photons turned out essential for quantum metrology, spectroscopy, imaging, and lithography \cite{Howard2019, bhusal2021smart, de_J_Le_n_Montiel_2013, Saleh05PRL, you2021scalable, obrien_photonic_2009}.

There has been important progress on the preparation of multiphoton systems with quantum mechanical properties \cite{magana2019multiphoton, DELLANNO200653}. The interest in these systems resides in the complex interference and scattering effects that they can host \cite{aspuru-guzik_photonic_2012, obrien_photonic_2009, You2020plasmonics}. Remarkably, these fundamental processes define the statistical fluctuations of photons that establish the nature of light sources \cite{Mandel:79, you2020identification,magana2019multiphoton, DELLANNO200653,mandel1995optical}. Furthermore, these quantum fluctuations are associated to distinct excitation modes of the electromagnetic field that determine the quantum coherence of a light field \cite{Mandel:79,mandel1995optical}. In the context of quantum information processing, the interference and scattering among photons have enormous potential to perform operations that are intractable on classical systems \cite{aspuru-guzik_photonic_2012,obrien_photonic_2009}. This feature of multiphoton systems has stimulated the development of optical circuits for quantum random walks and boson sampling to implement operations that result unfeasible for classical devices \cite{aspuru-guzik_photonic_2012,obrien_photonic_2009}. Despite recent progress demonstrated in quantum information processing with multiphoton systems, it is not possible to describe the evolution of their properties of quantum coherence. Indeed, the existing formulations of the van Cittert-Zernike theorem do not allow for the description of the quantum statistical properties of photonic systems \cite{Wolf:54, Dorrer:04, gori2000use, cai2020, Cai2012, Carozzi:09, Batarseh:18, Barakat:2000, Saleh05PRL, Fabre_2017, Howard2019, Barrachina2020, mandel1995optical}.

In this chapter, I introduce a quantum version of the van Cittert-Zernike theorem to describe the evolution of quantum coherence of propagating thermal multiphoton wavepackets. This work extends previous investigations that explored the classical evolution of spatial, temporal, spectral, and polarization coherence of optical fields \cite{Wolf:54, Dorrer:04, gori2000use, cai2020, Cai2012, Carozzi:09, Batarseh:18, Barakat:2000, Saleh05PRL, Fabre_2017, Howard2019, Barrachina2020, mandel1995optical}. My theory demonstrates that it is possible to exploit distinguishable and indistinguishable scattering among propagating photons to control their quantum properties of coherence \cite{you2021observation}. Interestingly, these interactions enable the preparation of multiphoton systems with sub-Poissonian statistics without non-linear interactions \cite{magana2019multiphoton, you2021observation, kondakci_photonic_2015}. Specifically, I show that the implementation of conditional measurements enables the preparation of multiphoton systems with attenuated quantum statistics below the shot-noise limit. As such, I believe that 
these findings, together with the multiphoton quantum van Cittert-Zernike theorem, will have important implications for the development of quantum technologies \cite{magana2019quantum,aspuru-guzik_photonic_2012,obrien_photonic_2009}.  This chapter is based on the work originally presented in Miller et al \cite{miller2022multiphoton}.
\section{Theory}
\begin{figure}[t!]
 \includegraphics[width=0.9\linewidth]{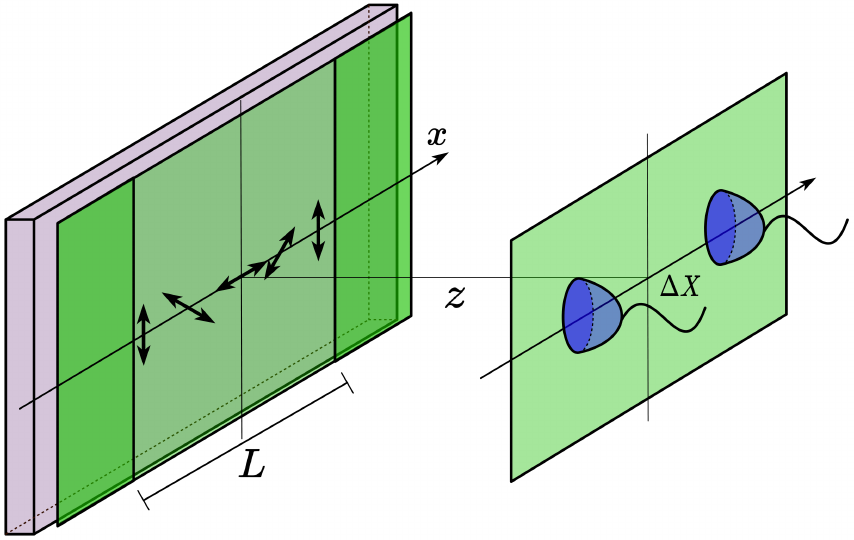}
 \caption[The proposed setup for investigating the multiphoton quantum van Cittert-Zernike theorem.]{The proposed setup for investigating the multiphoton quantum van Cittert-Zernike theorem. I consider an incoherent, unpolarized beam interacting with a polarization grating of length $L$ at $z=0$. After interacting with the grating, the beam propagates a distance of $z$ onto the measurement plane, where two point detectors  are placed $\Delta X$ apart.}
 \label{fig:Setup}
\end{figure}

I demonstrate the multiphoton quantum van-Cittert Zernike theorem by extending the work of Gori et al. \cite{gori2000use} to two-mode correlations using the setup in Fig. \ref{fig:Setup}. In general, each mode can host a multiphoton system with an arbitrary number of photons. I consider a thermal, spatially incoherent, unpolarized beam that interacts with a polarization grating. This grating modifies the polarization of the thermal beam at different transverse spatial locations $x$ according to ${\pi x}/{L}$. Here, $L$ represents the length of the grating. The thermal beam propagates to the far-field, where it is measured by two point detectors \cite{magana-loaiza-2016,Shih2009PRA,Chekhova08PRA}. I then post-select on the intensity measurements made by these two detectors to quantify the correlations between different modes of the beam.



The multiphoton quantum van Cittert-Zernike theorem can be demonstrated for any incoherent, unpolarized state, the simplest of which is an unpolarized two-mode state \cite{soderholm2001unpolarized}. The two-mode state can be produced by a source emitting a series of spatially independent photons with either horizontal (H) or vertical (V) polarization, giving an initial state \cite{magana2019multiphoton,wei2005synthesizing}
\begin{equation}
\begin{aligned}
\hat{\rho}&=\hat{\rho}_{1} \otimes \hat{\rho}_{2}\\
&= \frac{1}{4}\left(\left|H\right\rangle_{1}\left|H\right\rangle_{2}\left\langle H\right|_{1}\left\langle H\right|_{2}\right.+\left|H\right\rangle_{1}\left|V\right\rangle_{2}\left\langle H\right|_{1}\left\langle V\right|_{2} \\
&\quad+\left|V\right\rangle_{1}\left|H\right\rangle_{2}\left\langle V\right|_{1}\left\langle H\right|_{2}\left.+\left|V\right\rangle_{1}\left|V\right\rangle_{2}\left\langle V\right|_{1}\left\langle V\right|_{2}\right),
\end{aligned}
\label{eqn:state}
\end{equation}
where the subscripts denote the mode. For simplicity, I begin by considering the case where a single photon is emitted in each mode.


I find the state immediately after the polarization grating shown in Fig. \ref{fig:Setup} to be
\begin{equation}
\begin{aligned}
  \hat{\rho}_{\text{pol}}=\hat{P}\pare{x_1}\hat{\rho}_1\hat{P}\pare{x_2}\otimes\hat{P}\pare{x_3}\hat{\rho}_2\hat{P}\pare{x_4},
\end{aligned}
\label{eqn:pol}
\end{equation}
where $\hat{P}(x)$ is the projective measurement given by
\begin{equation}
\begin{aligned}
  \hat{P}\pare{x}=\begin{bmatrix}
  \cos^2\pare{\frac{\pi x}{L}} && \cos\pare{\frac{\pi x}{L}}\sin\pare{\frac{\pi x}{L}} \\
  \cos\pare{\frac{\pi x}{L}}\sin\pare{\frac{\pi x}{L}} &&\sin^2\pare{\frac{\pi x}{L}}
  \end{bmatrix}.
\end{aligned}
\label{eqn:Prop} 
\end{equation}

For ease of calculation, I utilize the Heisenberg picture, back-propagating the detector operators to the polarization grating. The point detector is modeled by  $\hat{O}_{j,k,z}(X)=\hat{a}^{\dag}_{j,z}(X)\hat{a}_{k,z}(X)$, where $z$ is the distance between the grating and the measurement plane. The ladder operator $\hat{a}_z(X)$ is defined as
\begin{equation}
\begin{aligned}
  \hat{a}_{j,z}(X)= \int^{\frac{L}{2}}_{-\frac{L}{2}} dx \hat{a}_{j,0}\pare{x}\text{Exp}[-\frac{2\pi i}{z\lambda}xX],
\end{aligned}
\label{eqn:ladOp}
\end{equation}
where $X$ is the position of the detector on the measurement plane, $\lambda$ is the wavelength of the beam and $j,k$ is the polarization of the operator. Eqn. (\ref{eqn:ladOp}) describes  the contribution of each point on the polarization grating plane to the detection measurement. Since I wish to keep the information of each interaction on the screen, I choose to calculate the four-point auto covariance by \cite{gureyev2017van, perez2017two}
\begin{equation}\label{eqn:Coh1}
\begin{aligned}
  G_{jklm}^{\pare{2}}&\pare{\mathbf{X},z}=\\&\text{Tr}[\hat{\rho}_{\text{pol}}\hat{a}^{\dag}_{j,z}\pare{X_1}\hat{a}_{k,z}\pare{X_2}\hat{a}^{\dag}_{l,z}\pare{X_3}\hat{a}_{m,z}\pare{X_4}],
\end{aligned}
\end{equation}
where $\mathbf{X}=\left[X_1,X_2,X_3,X_4\right]$, allowing for the measurement of a post-selected coherence. I then set $X_2=X_1$ and $X_4=X_3$, since I am working with two point detectors. I allow the operators of the two detectors to commute, recovering the well-known expression for second-order coherence \cite{gerry2005introductory}. The second-order coherence of any post-selected measurement is then found to be
\begin{equation}\label{eqn:2ndOrd}
\begin{aligned}
&G_{j k l m}^{(2)}(\mathbf{X},z)=\int d x_{1} \int d x_{2} \int d x_{3} \int d x_{4}\\
&\times C_{j k l m}(\boldsymbol{x}) F(\boldsymbol{x}, \mathbf{X},z) [\delta\left(x_{1}-x_{2}\right) \delta\left(x_{3}-x_{4}\right)\\
&+\delta\left(x_{1}-x_{4}\right) \delta\left(x_{3}-x_{2}\right)],
\end{aligned}
\end{equation}
where the limits of integration for each integral is $-L/2$ to $L/2$, $\boldsymbol{x}=[x_1,x_2,x_3,x_4]$, $C_{jklm}\pare{\bm{x}}$ is the coefficient of the $\ket{j}_1\ket{k}_2\bra{l}_1\bra{m}_2$ element of the density matrix $\hat{\rho}_{\text{pol}}$ in Eqn. (\ref{eqn:pol}), and $j,k,l,m \in \{H,V\}$. Furthermore, $F\pare{\bm{x},\mathbf{X},z}$ is given as
\begin{equation}
\begin{aligned}
  F\pare{\bm{x},\mathbf{X},z}=\text{Exp}[\frac{2\pi i}{\lambda z}\pare{X_4x_4-X_3x_3+X_2x_2-X_1x_1}].
\end{aligned}
\label{eqn:prop}
\end{equation}
\begin{figure}
 \includegraphics[width=0.95\linewidth]{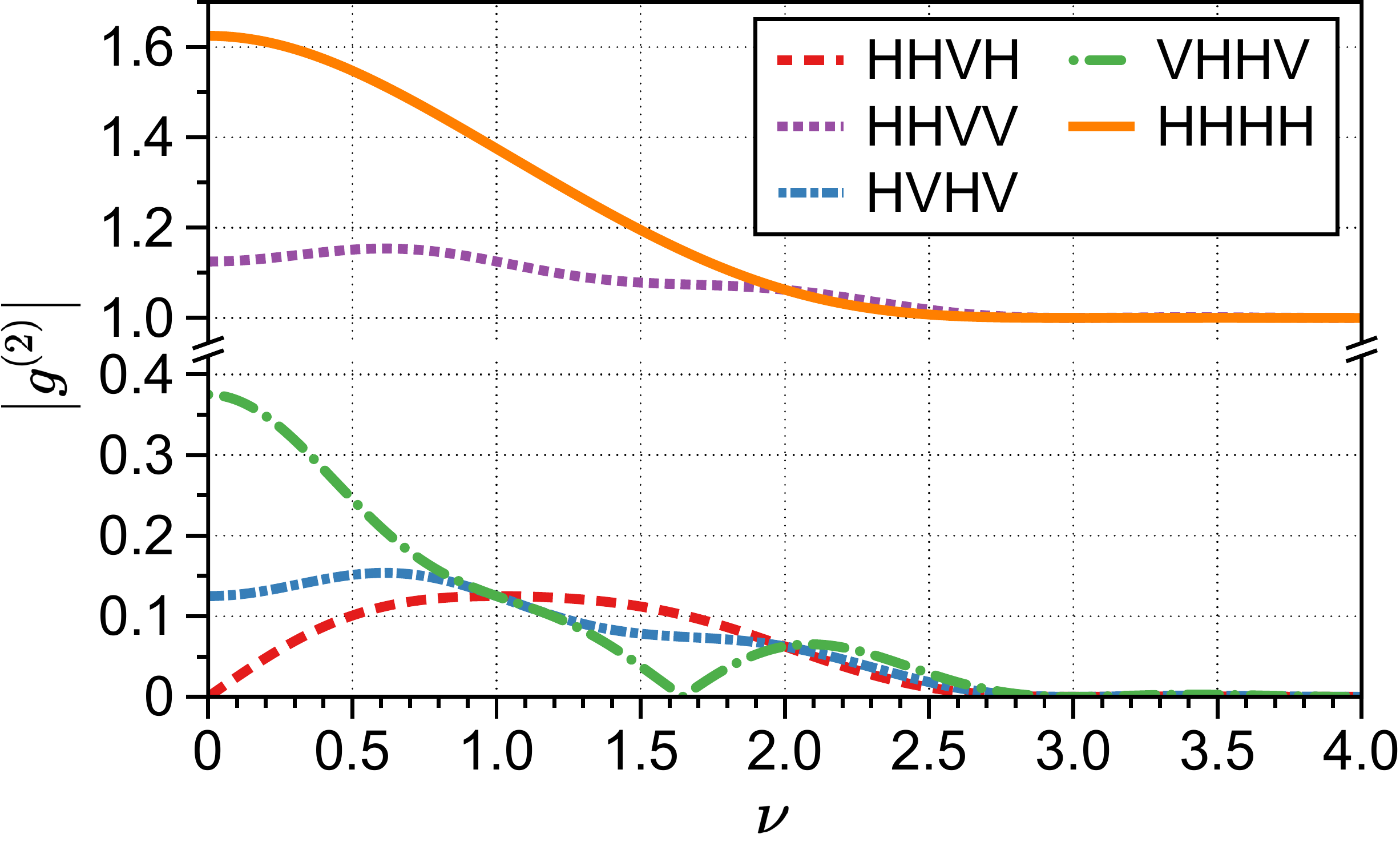}
 \caption[The second-order coherence for various post-selected measurements.]{The second-order coherence for various post-selected measurements. The $x$-axis is how the $g^{(2)}$ changes as a function of $\nu=L\Delta X/\pare{\lambda z}$ while keeping $L$, $\lambda$ and $z$ fixed. As the detectors move further apart, the spatial correlations created by the polarization grating decrease until they diminish entirely at $\nu\approx2.7$. In addition, certain post-selected measurements allow me to quantify the coherence between two fields that possess sub-Poissonian statistics, suggesting the possibility of sub-shot noise measurements. Note, these measurements can be also performed using quantum state tomography.} 
 \label{fig:Components}
\end{figure}
I set $X_2=X_1$ and $X_4=X_3$, which properly describes the two point detectors allowing Eqn. (\ref{eqn:2ndOrd}) to become a 2D Fourier transform \cite{gori2000use}. By observing  Eqn. (\ref{eqn:2ndOrd}), it is important to note that there are two spatial correlations that contribute to the coherence at the measurement plane. One is the correlation of a photon with itself which existed prior to interacting with the polarizer, while the other is the spatial correlation gained between the two photons upon interaction with the polarization grating. Due to the nature of projective measurements in Eqn. (\ref{eqn:pol}), the density matrix $\hat{\rho}_{\text{pol}}$ will no longer be diagonal in the horizontal-vertical basis, allowing for the beam to temporarily gain and lose polarization coherence \cite{gori2000use}. The self-coherence of a photon results in the minimum coherence throughout all measurements in the far-field. The correlations between different photons sets the maximum coherence and determines how it changes with the distance between the detectors.

To extend the description of a two-mode system comprising of two photons, I need to move beyond a purely quantum picture. Attempting to propagate a multiphoton field under the Schrodinger and Heisenberg pictures becomes computationally hard, scaling on the order of $O(2^n n!)$ where $n$ represents the number of photons \cite{10.1145/1993636.1993682}. As a result, this new formalism describes the evolution of  multiphoton systems, using the beam coherence-polarization (BCP) matrix \cite{gori1998beam,gori2000use}
\begin{align}
\ave{\hat{J}\pare{X_1,X_2,z}}=
\begin{bmatrix}
\ave{\hat{E}_H^\dag\pare{X_1,z}\hat{E}_H\pare{X_2,z}} && \ave{\hat{E}_H^\dag\pare{X_1,z}\hat{E}_V\pare{X_2,z}}\\
\ave{\hat{E}_V^\dag\pare{X_1,z}\hat{E}_H\pare{X_2,z}} && \ave{\hat{E}_V^\dag\pare{X_1,z}\hat{E}_V\pare{X_2,z}}
\end{bmatrix}.
\end{align}
Here, the angle brackets denote time average, whereas the quantities $\hat{E}_{\alpha}^\dag\pare{X,z}$ and $\hat{E}_{\alpha}\pare{X,z}$ represent the negative- and positive-frequency components of the $\alpha$-polarized (with $\alpha = H,V$) field-operator at the space-time point $(X,z;t)$, respectively. We can then propagate the BCP matrix through the grating and to the measurement plane, by considering an initial BCP matrix of the form: $I_2\delta\pare{X_1-X_2}$ \cite{gori1998beam}. The details of the calculation of propagation can be found in Appendix \ref{app2}. Upon reaching the measurement plane we can find the second-order coherence matrix given by \cite{Pires:21}
\begin{equation} \label{eqn:Coh2}
\begin{aligned}
 \mathbf{G}^{(2)}(\bm{X},z)=\ave{\hat{J}\pare{X_1,X_2,z}\otimes\hat{J}\pare{X_3,X_4,z}}.
\end{aligned}
\end{equation}
\\
Each element of the $\mathbf{G}^{(2)}$ matrix is a post-selected coherence matching each combination of polarizations shown in Eqns. (\ref{eqn:Coh1})-(\ref{eqn:2ndOrd}). As shown in Appendix \ref{app2}, the result obtained is equivalent to the approach described in Eqns. (\ref{eqn:state})-(\ref{eqn:prop}).
\section{Analysis and Applications}

In order to demonstrate the results of our calculation, we first look at the second-order coherence of the horizontal mode in the far-field. By normalizing either Eqn. (\ref{eqn:Coh1}) or the matrix element of Eqn. (\ref{eqn:Coh2}), we find the coherence of the horizontal mode to be
\begin{equation}
\begin{aligned}
  g^{(2)}_{\text{HHHH}}&(\nu)=1+\frac{1}{16}\sinc^2\pare{2-\nu}+\frac{5}{8}\sinc^2\pare{\nu}+\frac{1}{16}\sinc^2\pare{2+\nu}\nonumber+\frac{1}{4}\sinc\pare{2-\nu}\sinc\pare{1-\nu}\nonumber\\&+\frac{3}{8}\sinc^2\pare{1-\nu}
  +\frac{1}{4}\sinc\pare{1+\nu}\pare{\sinc\pare{2+\nu}+\sinc\pare{1-\nu}}+\frac{3}{8}\sinc^2\pare{1+\nu}\nonumber\\
  &+\frac{1}{8}\sinc\pare{\nu}\pare{\sinc\pare{2-\nu}+\sinc\pare{2+\nu}+6\sinc\pare{1-\nu}+6\sinc\pare{1+\nu}},
\end{aligned}
\label{eqn:g2HHHH}
\end{equation}
where $g^{(2)}_{jklm}$ is the normalized second-order coherence. Here $\sinc(\nu)=\sin(\pi \nu)/\pare{\pi \nu}$ and $\nu=L\Delta X/\pare{\lambda z}$. Therefore, $g^{(2)}_{jklm}$ depends on the distance between the detectors $\Delta X = X_1 - X_2$, the length of the polarization grating $L$, the wavelength $\lambda$, and the distance in the far field $z$. The same holds true for all other ${g}^{(2)}_{jklm}$, where each expression can be found in Appendix \ref{app2}. Since Eqn. (\ref{eqn:g2HHHH}) applies to all incoherent unpolarized states we will perform the analysis for two mode thermal states, as the statistical properties are well studied \cite{mandel1995optical,gerry2005introductory,you2021observation,you2020identification,you2021observation}.
\begin{figure*}
 \includegraphics[width=0.95\linewidth]{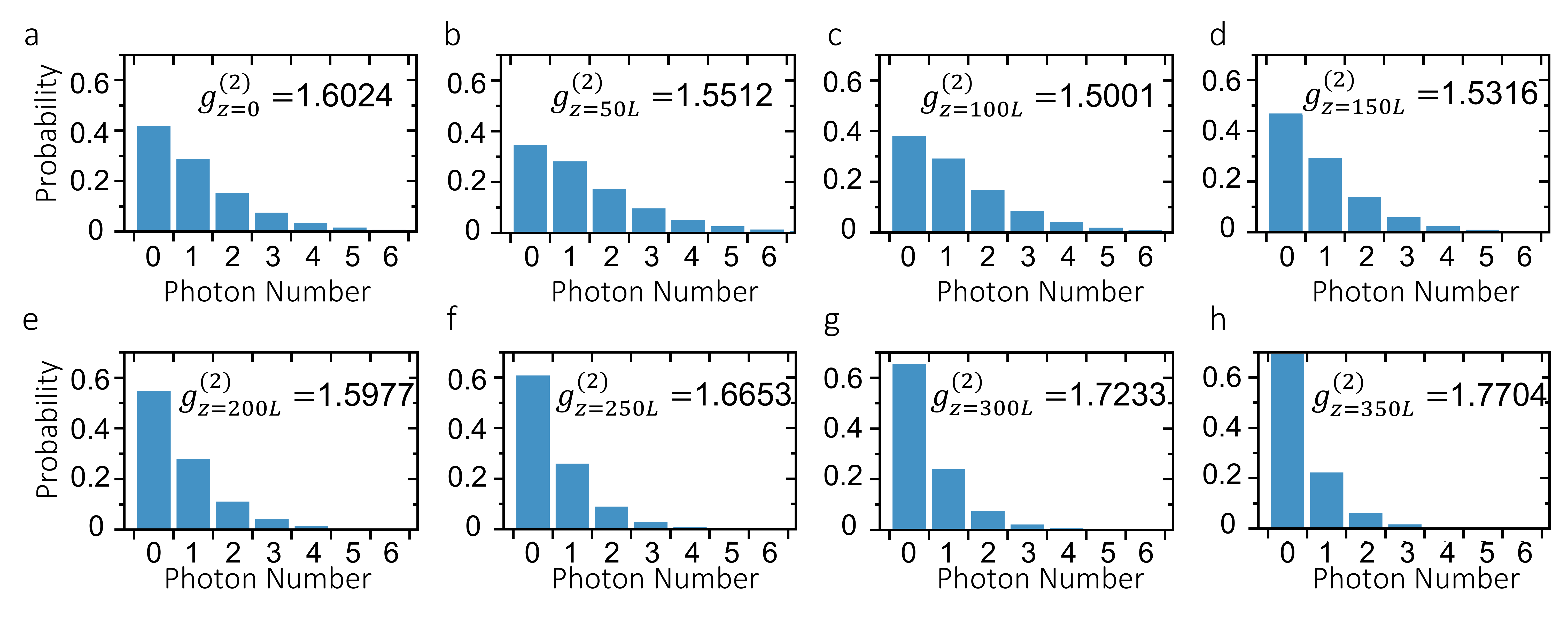}
 \caption[The modification of the photon-number distribution and quantum coherence of a thermal multiphoton system upon propagation.]{ The modification of the photon-number distribution and quantum coherence of a thermal multiphoton system upon propagation. In this case, the multiphoton system comprises a mixture of single-mode photons with either vertical or horizontal polarization. I assumed a single photon-number-resolving detector placed at different propagation distances: \textbf{a} $z=0$, \textbf{b} $z=50L$, \textbf{c} $z=100L$, \textbf{d} $z=150L$, \textbf{e} $z=200L$, \textbf{f} $z=250L$, \textbf{h} $z=300L$, \textbf{f} $z=350L$. In the transverse plane, the photon-number-resolving detector is placed at $X=0.4 L$. } 
 \label{fig:Components2}
\end{figure*}

As shown in Fig. \ref{fig:Components}, increasing the separation $\Delta X$ of the detectors causes the correlations to gradually decrease. Once $\nu\approx 2.7$, the detectors become uncorrelated. Noting that $g^{(2)}(\nu)=1, \nu\neq 0$ represents an uncorrelated measurement since this can only be true when the two spatial modes become separable. In addition, when one of the two measured modes is no longer contributing to the measurement we get a $g^{(2)}(\nu)=0$. Interestingly, by fixing the distance $\Delta X$ between the two detectors, I can increase the correlations by moving the measurement plane further into the far-field. This is equivalent to decreasing $\nu$, causing correlations to increase to a possible maximum value of $g^{(2)}(0)=1.62$, suggesting an increase in bunching \cite{mandel1995optical}. By measuring $g^{(2)}_{\text{grating}}(0)$ immediately after the polarization grating at $x=0$, a horizontally polarized beam is measured with a $g^{(2)}_{\text{grating}}(0)=2$. Noting the theory we presented only applies to the far-field, therefore these two values do not contradict each other. While the exact transition between the near and the far-fields are beyond the scope of the paper, we note that the horizontal mode along the central axis becomes more coherent as it propagates to the far-field, as predicted by the van Cittert-Zernike theorem \cite{gureyev2017van, Fabre_2017}. 

Setting one detector to measure the vertical mode and the other detector the horizontal mode, given by $g^{(2)}_{\text{HHVV}}$ in Fig. \ref{fig:Components}, I can measure the coherence between the horizontal and vertical mode. This post-selective measurement results in a different effect from when we only measured only the horizontal mode. Placing the detectors immediately after the polarization grating at $x=0$, I measure $g^{(2)}_{\text{grating}}(0)=0$ since there is no vertically polarized mode. However, I measure $g^{(2)}(0)\approx 1.1$ when the beam is propagated into the far-field. In this case, the polarization grating leads to the thermalization of the beam \cite{you2021observation}. This can be verified by removing the polarization grating and repeating the measurement giving $g^{(2)}_{\text{initial}}(0)=1$ since the two modes are completely uncorrelated.

The measurements of $g^{(2)}_{\text{HHHH}}$ and $g^{(2)}_{\text{HHVV}}$ can be performed using point detectors, however I predict more interesting effects that can be observed through the full characterization of the field. This information can be obtained through quantum state tomography \cite{cramer2010efficient}. I find that the second-order coherence $g^{(2)}_{\text{VHHV}}$, $g^{(2)}_{\text{HHVH}}$ and $g^{(2)}_{\text{HVHV}}$ is below one suggesting sub-Poissonian statistics, which potentially allows for sub-shot-noise measurement \cite{agarwal2012quantum}. It is important to note that while $g^{(2)}_{\text{HHVH}}(0)<g^{(2)}_{\text{HHVH}}(\nu)$  would be indicative of anti-bunching \cite{agarwal2012quantum}, however the term is imaginary. This feature is found using the BCP matrix approach, therefore it is true for all unpolarized incoherent fields. The sub-Poissonian statistics were achieved only with the use of post-selection without nonlinear interactions \cite{magana2019multiphoton,you2021scalable,obrien_photonic_2009}. Another interesting feature is that the $g^{(2)}_{\text{VHHV}}$ decays and resurrects  at $\nu\approx 1.6$ before decaying again. 

The sub-Possonian statistics are exclusive to unpolarized systems.  Returning to Eqn. (\ref{eqn:2ndOrd}) note that there are two correlations contributing to the final coherence, one from the photons self-coherence that existed prior to interaction with the screen and another coherence term that comes from the interaction.  For unpolarized states there is no initial correlations in the off-diagonal elements of the density matrix, since by definition the off diagonal elements are zero \cite{soderholm2001unpolarized}.  This results in the first term of Eqn. (\ref{eqn:2ndOrd}) to be zero for all diagonal elements.  As noted above, this term sets the minimum value of the coherence measurement to zero, allowing for the measurement of sub-Poissonian statistics. 

Finally, I would like to highlight the fact that the quantum statistical properties of multiphoton systems can change upon propagation
due to the individual interactions of their constituent single-mode photons carrying different polarizations. This effect is quantified through the second-order quantum coherence $g^{(2)}(\tau=0)$ defined as $g^{(2)}(\tau=0)=1+\left(\left\langle(\Delta \hat{n})^{2}\right\rangle-\langle\hat{n}\rangle\right) /\langle\hat{n}\rangle^{2}$ \cite{Mandel:79, mandel1995optical}. In this case, the averaged quantities in $g^{(2)}(\tau=0)$ are obtained through the density matrix of the system's state, as described in Eqn. (\ref{eqn:state}),
at different spatial coordinates ($\boldsymbol{X},z$). In Figure \ref{fig:Components2}, we report the photon-number distribution of the combined vertical-horizontal multiphoton field. In this case, a single photon-number-resolving detector was placed at $X=0.4 L$ \cite{you2020identification}. Note that by selecting the proper propagation distance $z$, one could, in principle, generate on-demand multiphoton systems with sub-Poissonian or Poissonian statistics \cite{magana2019multiphoton,you2021observation}. As indicated in Figs. \ref{fig:Components} and  \ref{fig:Components2}, the evolution of quantum coherence  upon propagation lies at the heart of the quantum van Cittert-Zernike theorem for multiphoton systems.
\section{Conclusion}
In conclusion, I have investigated new mechanisms to control nonclassical coherence of multiphoton systems. I describe these interactions using a quantum version of the van Cittert-Zernike theorem. Specifically, by considering a polarization grating together with conditional measurements, I show that it is possible to control the quantum coherence of multiphoton systems. Moreover, I demonstrate the possibility of producing multiphoton systems with
sub-Poissonian statistics through linear interactions \cite{you2021observation,DELLANNO200653,kondakci_photonic_2015}. My work demonstrates that the multiphoton quantum van Cittert-Zernike theorem will have important implications for describing the evolution of the properties of quantum coherence of many-body bosonic systems \cite{You2020plasmonics,DELLANNO200653}.

\chapter*{Chapter 6. \\ Concluding Remarks}
\addcontentsline{toc}{chapter}{Chapter 6. Concluding Remarks}
\setcounter{section}{0}
\setcounter{figure}{0}
\setcounter{table}{0}
\refstepcounter{chapter}
\label{Conclusion}

\doublespacing
This dissertation began with a summary of current work in quantum technologies focusing on quantum information processing, computing and sensing, establishing the importance of quantum coherence to the advancement of quantum technologies.  In Chapter \ref{Intro}, I presented a brief background in quantum optics, machine learning, and the classical limits that would be investigated in this disseration.  In Chapter \ref{Coherence}, I proceeded with a more advanced background, this time focused only on coherent phenomena both in classical optics as well as quantum theories.  Throughout chapters \ref{Quantum Networks}, \ref{QI} and \ref{MQvCZT} I provided a summary on my work on coherence that was completed during by term at LSU. Here in the final chapter I will summarize and tie together my main results in order to provide a clear picture its potential impact.

The advancement of quantum technologies relies on a complete understanding of quantum coherence.  In Chapter \ref{Quantum Networks}, I demonstrated the power of this knowledge by designing a quantum network with a simple decoherence model.  I then applied this model to common operation performed in quantum networking, namely entanglement swapping and purification, to build an algebra.  This algebra in turn provided a set of rules demonstrating how to perform a graph reduction on the network. Applying these rules allowed the derivation of basic principles of quantum network design and routing, proving that the shortest path method in classical networking is insufficient in quantum networking.  Instead, quantum networking is a graph partitioning problem, where the goal is to find the optimal subgraph between two users.  This, in turn, enabled the design of a rudimentary two user routing algorithm.  This provides a clear picture of how refined knowledge of coherence can be exploited to reduce complicated problems. Furthermore, this work states the key considerations for routing and network design for the future development of a quantum internet \cite{rohde2021quantum,khatri_spooky_2021,PhysRevA.97.012335}.

In Chapter \ref{QI}, I derived a general model capable of describing the photon statistics for any single or multi-mode combination of thermal and coherent sources.  This model was then used to design a neural network capable of distinguishing the sources on a single pixel. This neural network was then applied to a single camera with different sources, consisting of two thermal and one coherent source, incident on it.  Here it was shown that individual picture on the image were able to distinguish the number of incident sources, even when completely overlapped. Then a two source simulation and experiment were performed.  During this both the experimental and theoretical ability to resolve two sources were resolve the two sources by our scheme, as well as classical means, were found.  I then showed that our scheme surpasses the Rayleigh-Abbe criterion achieving super-resolution.  This work represents an important step in building a quantum LIDAR as well as having an important impact to Astronomy \cite{doi:10.1063/1.4770298,doi:10.1063/1.4809836}.  Our scheme, unlike previous schemes, only requires knowledge of the type of light involved in order to properly adjust the neural network \cite{magana2019quantum, won2009eyes, stelzer2002beyond, kolobov2000quantum, stelzer2000uncertainty,editorial2009, pirandola2018advances}.  This gives it a unique advantage over previous schemes that relied on strict knowledge of the spatial location of the incoming beams, which can prevent the schemes from being applied in situation where strict control of the source is unavailable \cite{tsang2016quantum,Steinberg17PRL,zhou2019quantum,Treps20Optica,Saleh18Optica,Liang21Optica}. 

Lastly, in Chapter \ref{MQvCZT}, I demonstrated a quantum version of the van Cittert-Zernike theorem capable of modeling the propagation of coherence properties for multiphoton systems.  This model was then applied to analyze the coherence properties of an unpolarized coherent beam when acted on by a polarization grating.  I demonstrated that in the far field, certain post-selected measurements are capable of producing sub-Poissonian statistics and anti-bunching, without the use of a nonlinear media \cite{you2021observation,DELLANNO200653,kondakci_photonic_2015}.  Furthermore, I then analyzed the field under free propagation, and showed that changes in coherence are directly related to changes in photon statistics.  This work unlocks key insights into the propagation of quantum coherence. As shown through the generation of states with sub-Poissonian states, this formalism opens up new ways to describe the evolution of many bosonic systems \cite{You2020plasmonics,DELLANNO200653}. Which is essential for the implementation of quantum devices.

As the research currently stands, these three works stand separate with each one providing useful insights into quantum coherence, but with the effects of each one independent of the other. However,  the creation of the multiphoton quantum van Cittert-Zernike theorem enables more in depth and accurate analysis of quantum systems.  In the case of imaging, we can potentially use our newly found understanding around the propagation of coherence properties to improve our current imaging model through the application of grey box machine learning techniques, which has shown promise in quantum control problems for qubits \cite{Perdomo_Ortiz_2018,Perrier_2020,bhusal2021spatial,lollie2021highdimensional}.  In terms of quantum networks, an improved knowledge of how a state propagates gives the potential for improved model of decoherence.  This gives the potential introduce better error models and therefore algebra's to improve our ability to optimize network routing. Furthermore, the potential of the quantum van Cittert-Zernike theorem to state preparation presents the potential for new quantum network channels capable of transmitting data at higher rates, another problem currently limiting quantum networks \cite{leone2021qunet,rohde2021quantum}. 
%
\appendix
%
\singlespacing
\chapter*{Appendix A. \\ Details for Quantum Imaging}
\addcontentsline{toc}{chapter}{Appendix A.  Details for Quantum Imaging}
\setcounter{chapter}{1}
\setcounter{section}{0}
\setcounter{figure}{0}
\setcounter{table}{0}
\label{app1}
\doublespacing
\section{Methods}\label{app1:Methods}
\subsection{Training of NN}
For the sake of simplicity, I split the functionality of the neural network  into two phases: the training and testing phase. In the first phase, the training data is fed to the network multiple times to optimize the synaptic weights through a scaled conjugate gradient back-propagation algorithm \cite{moller1993scaled}. This optimization seeks to minimize the Kullback-Leibler divergence distance between predicted and the real target classes \cite{kullback1951information, kullback1997information}. At this point, the training is stopped if the loss function does not decrease within 1000 epochs \cite{prechelt1998early}. In the test phase, I assess the performance of the algorithm by introducing an unknown set of data during the training process. For both phases,  I prepare a data-set consisting of one thousand experimental measurements of photon statistics for each of the five classes. This process is formalized by considering different numbers of data points: 100, 500, ..., 9500, 10000. Following a standardized ratio for statistical learning, I divide the data into training (70\%), validation (15\%), and testing (15\%) sets \cite{crowther2005method}.  The networks were trained using the neural network toolbox in MATLAB, which runs on a computer Intel Core i7–4710MQ CPU (@2.50GHz) with 32GB of RAM.
\subsection{Fittings}
To determine the optimal fits for Fig. \ref{profiles}\textbf{e} and \textbf{f} I design a search space based on Eqs. \ref{pthcoh} and \ref{pthcohdis}.  To do so I first found the mean photon number of the input pixel, which will later be applied to constrain the search space.  From here I allowed for the existence of up to three distinguishable modes which will be combined according to Eq. \ref{pthcohdis}.  Each of the modes contains an indistinguishable combination of up to one coherent and two thermal sources whose number distribution is given by Eq. \ref{pthcoh}.  The total combination results in partially distinguishable combination and provides the theoretical model for our experiment.  From here the search space is
\begin{align*}
    \sqrt{\sum_{n=0}(p_{\text{exp}}(n)-p_{\text{th}}(n|\vec{n}_{1,t},\vec{n}_{2,t},\vec{n}_c))^2},
\end{align*}  
where $\vec{n}_{i,t}$ and $\vec{n}_{c}$ are the mean photon numbers of that each thermal or coherent source contributes to each distinguishable mode respectively.  The mean photon numbers of each source must add up to the experimental mean photon number, constraining the search.  A linear search was then performed over the predicted mean photon numbers and the minimum was returned, providing the optimal fit.

\subsection{Monte-Carlo Simulation of the Experiment}
To demonstrate a consistent improvement over traditional methods, I also simulated the experiment using two beams, a thermal and a coherent, with Gaussian point spread functions over a 128$\times$128 grid of pixels. At each pixel, the mean photon number for each source is provided by the Gaussian point spread function, which is then used to create the appropriate distinguishable probability distribution as given in Eq. \ref{pthcohdis}, creating a 128$\times$128 grid of photon number distributions. The associated class data for these distributions will then be fitted using to a set of pre-labeled disks using a genetic algorithm.  This recreates the method in the limits of perfect classification. Each of these distributions is then used to simulate photon-number resolving detection. This data is then used to create a normalized intensity for the classical fit. I fit the image to a combination of Gaussian PSFs. This process is repeated ten times for each separation in order to average out fluctuations in the fitting. When combining the results of the intensity fits they are first divided into two sets.  One set has the majority of fits return a single Gaussian, while the other returned two Gaussian the majority of the time. The set identified as only containing a single Gaussian is then set at the Abbe-Rayleigh diffraction limit, while the remaining data is used in a linear fit. This causes the sharp transition between the two sets of data.
\section{Derivation of the Many-Source Photon-Number Distribution}\label{app1:Derivation}
Start by considering the indistinguishable detection of $N$ coherent and $M$ thermal independent sources. To obtain the combined photon distribution, I make use of the Glauber-Sudarshan theory of coherence \cite{glauber1963quantum,sudarshan1963equivalence}. Thus, I start by writing the P-functions associated to the fields produced by the indistinguishable coherent and thermal sources, which is,
\begin{equation}\label{Eq:Pcoh}
P_{\text{coh}}\pare{\alpha} = \int P^{\text{coh}}_{N}\pare{\alpha - \alpha_{_{N-1}}}P^{\text{coh}}_{N-1}\pare{\alpha_{_{N-1}}-\alpha_{_{N-2}}}\cdots P^{\text{coh}}_{2}\pare{\alpha_{_{2}}-\alpha_{_{1}}}P^{\text{coh}}_{1}\pare{\alpha_{_{1}}}d^{2}\alpha_{_{N-1}}\cdots d^{2}\alpha_{_{1}},
\end{equation}
\begin{equation}\label{Eq:Pth}
P_{\text{th}}\pare{\alpha} = \int P^{\text{th}}_{M}\pare{\alpha - \alpha_{_{M-1}}}P^{\text{th}}_{M-1}\pare{\alpha_{_{M-1}}-\alpha_{_{M-2}}}\cdots P^{\text{th}}_{2}\pare{\alpha_{_{2}}-\alpha_{_{1}}}P^{\text{th}}_{1}\pare{\alpha_{_{1}}}d^{2}\alpha_{_{M-1}}\cdots d^{2}\alpha_{_{1}},
\end{equation}
with $P_{\text{coh}}\pare{\alpha}$ and $P_{\text{th}}\pare{\alpha}$ standing for the P-functions of the combined $N$-coherent and $M$-thermal sources, respectively. In both equations, $\alpha$ stands for the complex amplitude as defined for coherent states $\ket{\alpha}$, and the individual-source P-functions are defined as
\begin{equation}\label{Eq:P1coh}
P^{\text{coh}}_{k}\pare{\alpha} = \delta^{2}\pare{\alpha - \alpha_{k}},
\end{equation}
\begin{equation}\label{Eq:P1th}
P^{\text{th}}_{l}\pare{\alpha} = \frac{1}{\pi\bar{m}_{l}}\exp\pare{-\abs{\alpha}^{2}/\bar{m}_{l}},
\end{equation}
where $P^{\text{coh}}_{k}\pare{\alpha}$ corresponds to the P-function of $k$th coherent source, with mean photon number $\bar{n}_{k} = \abs{\alpha_{k}}^2$, and $P^{\text{th}}_{l}\pare{\alpha}$ describes the $l$th thermal source, with mean photon number $\bar{m}_{l}$.
\\
Now, by substituting Eq. (\ref{Eq:P1coh}) into Eq. (\ref{Eq:Pcoh}), and Eq. (\ref{Eq:P1th}) into Eq. (\ref{Eq:Pth}), I obtain
\begin{equation}
P_{\text{coh}}\pare{\alpha} = \delta^{2}\pare{\alpha - \sum_{k=1}^{N}\alpha_{k}},
\end{equation}
\begin{equation}
P_{\text{th}}\pare{\alpha} = \frac{1}{\pi\sum_{l=1}^{M}\bar{m}_{l}}\exp\pare{-\frac{\abs{\alpha}^{2}}{\sum_{l=1}^{M}\bar{m}_{l}}}.
\end{equation}
\\
I can finally combine the thermal and coherent sources by writing
\begin{equation}\label{Eq:Pth-coh}
P_{\text{th-coh}}\pare{\alpha} = \int P_{\text{th}}\pare{\alpha - \alpha'}P_{\text{coh}}\pare{\alpha'}d^2\alpha'.
\end{equation}
Note that this expression enables the analytical description for the photon-number distribution $p_{\text{th-coh}}\pare{n}$ of an arbitrary number of indistinguishable sources measured by a quantum detector. Also notice that Eq. (\ref{Eq:Pth-coh}) is equivalent to Eq. (1) in the main text. More specifically, we can write
\begin{equation}\label{Eq:photon_th_coh}
p_{\text{th-coh}}\pare{n} = \bra{n}\hat{\rho}_{\text{th-coh}}\ket{n},
\end{equation}
where
\begin{equation}\label{Eq:rho}
\hat{\rho}_{\text{th-coh}} = \int P_{\text{th-coh}}\pare{\alpha}\ket{\alpha}\bra{\alpha}d^{2}\alpha,
\end{equation}
describes the the density matrix of the quantum states of the combined thermal-coherent field at the quantum detector.

Thus, by substituting Eq. (\ref{Eq:Pth-coh}) into (\ref{Eq:rho}) and (\ref{Eq:photon_th_coh}), I find that the photon distribution of the combined fields is given by
\begin{equation}\label{pthcoh}
\begin{split}
p_{\text{th-coh}}(n)=&\frac{\left(m_{\text{tot}}\right)^{n} \exp \left(-\abs{\alpha_{\text{tot}}}^{2} / m_{\text{tot}}\right)}{\pi\left(m_{\text{tot}}+1\right)^{n+1}}\sum_{k=0}^{n} \frac{1}{k !(n-k) !} \Gamma\left(\frac{1}{2}+n-k\right) \Gamma\left(\frac{1}{2}+k\right)\\
& \times {}_1F_{1}\left(\frac{1}{2}+n-k; \frac{1}{2}; \frac{(\operatorname{Re}[\alpha_{\text{tot}}])^{2}}{m_{\text{tot}}\left(m_{\text{tot}}+1\right)}\right){}_1F_{1}\left(\frac{1}{2}+k;\frac{1}{2}; \frac{(\operatorname{Im}[\alpha_{\text{tot}}])^{2}}{m_{\text{tot}}\left(m_{\text{tot}}+1\right)}\right),
\end{split}
\end{equation}
with $m_{\text{tot}} = \sum_{l=1}^{M}\bar{m}_{l}$ and $\alpha_{\text{tot}} = \sum_{k=1}^{N}\alpha_{k}$. In this final result, which corresponds to Eq. (2) of the main text, $\Gamma(z)$ and ${}_1F_1(a;b;z)$ are the Euler gamma and the Kummer confluent hypergeometric functions, respectively.
\singlespacing
\chapter*{Appendix B. \\ Multiphoton Quantum Van Cittert-Zernike Theorem via the Beam Coherence Polarization Matrix}
\addcontentsline{toc}{chapter}{Appendix B.  Multiphoton Quantum Van Cittert-Zernike Theorem via the Beam Coherence Polarization Matrix}
\refstepcounter{chapter}
\setcounter{section}{0}
\setcounter{figure}{0}
\setcounter{table}{0}
\label{app2}
\doublespacing
\noindent Start by considering the second-order coherence matrix for a polarized, quasi-monochromatic field \cite{gori2000use}
\begin{equation}\label{Eq:matrix1}
J\left(\boldsymbol{r}_{1}, \boldsymbol{r}_{2}, z\right)=\left\langle j\left(\boldsymbol{r}_{1}, \boldsymbol{r}_{2}, z\right)\right\rangle=\left[\begin{array}{ll}
J_{H H}\left(\boldsymbol{r}_{1}, \boldsymbol{r}_{2}, z\right) & J_{H V}\left(\boldsymbol{r}_{1}, \boldsymbol{r}_{2}, z\right) \\
J_{V H}\left(\boldsymbol{r}_{1}, \boldsymbol{r}_{2}, z\right) & J_{V V}\left(\boldsymbol{r}_{1}, \boldsymbol{r}_{2}, z\right)
\end{array}\right],
\end{equation}
where $z$ stands for the propagation distance of the field and $\boldsymbol{r}$ is used to specify the position of a point at the transverse plane of observation. The elements of the coherence-polarization matrix [Eq. (\ref{Eq:matrix1})] are given by
\begin{equation}
J_{\alpha \beta}\left(\boldsymbol{r}_{1}, \boldsymbol{r}_{2}, z\right)=\left\langle j_{\alpha \beta}\left(\boldsymbol{r}_{1}, \boldsymbol{r}_{2}, z\right)\right\rangle=\left\langle E_{\alpha}^{(-)}\left(\boldsymbol{r}_{1}, z ; t\right) E_{\beta}^{(+)}\left(\boldsymbol{r}_{2}, z ; t\right)\right\rangle,
\end{equation}
with $\alpha,\beta = H,V$. The angle brackets denote time average, whereas the quantities $E^{(-)}_{\alpha}\pare{\boldsymbol{r},z;t}$ and $E^{(+)}_{\alpha}\pare{\boldsymbol{r},z;t}$ represent the negative- and positive-frequency components of the $\alpha$-polarized field-operator at the space-time point $\pare{\boldsymbol{r},z;t}$, respectively.

Note that Eq. (\ref{Eq:matrix1}) is valid for classical fields, as well as single-photon sources \cite{gori2000use,Saleh05PRL}. However, to include multi-photon effects higher-order correlation functions are needed. In particular, for light in an arbitrary quantum state, the polarized, two-photon four-point correlation matrix can readily be written as
\begin{equation}\label{Eq:matrix2}
G\left(\boldsymbol{r}_{1}, \boldsymbol{r}_{2} ; \boldsymbol{r}_{3}, \boldsymbol{r}_{4}, z\right)=\left\langle j\left(\boldsymbol{r}_{1}, \boldsymbol{r}_{2}, z\right) \otimes j\left(\boldsymbol{r}_{3}, \boldsymbol{r}_{4}, z\right)\right\rangle,
\end{equation}
where $\otimes$ stands for the Kronecker (tensor) product. Note that the elements defined by the matrix in Eq. (\ref{Eq:matrix2}) are given by the four-point correlation matrix \cite{perez2017two}
\begin{equation}
G_{\alpha \beta \alpha^{\prime} \beta^{\prime}}\left(\boldsymbol{r}_{1}, \boldsymbol{r}_{2} ; \boldsymbol{r}_{3}, \boldsymbol{r}_{4}, z\right)=\left\langle E_{\alpha}^{(-)}\left(\boldsymbol{r}_{1}, z ; t\right) E_{\beta}^{(+)}\left(\boldsymbol{r}_{2}, z ; t\right) E_{\alpha^{\prime}}^{(-)}\left(\boldsymbol{r}_{3}, z ; t\right) E_{\beta^{\prime}}^{(+)}\left(\boldsymbol{r}_{4}, z ; t\right)\right\rangle,
\end{equation}
with $\alpha,\beta, \alpha', \beta' = H,V$. Furthermore, realize that the elements defined by the previous equation follow a propagation formula of the form \cite{gori1999}
\begin{equation}\label{Eq:twophoton}
\begin{gathered}
G_{\alpha \beta \alpha^{\prime} \beta^{\prime}}\left(\boldsymbol{r}_{1}, \boldsymbol{r}_{2} ; \boldsymbol{r}_{3}, \boldsymbol{r}_{4}, z\right)=\int\int\int\int G_{\alpha \beta \alpha^{\prime} \beta^{\prime}}\left(\boldsymbol{\rho}_{1}, \boldsymbol{\rho}_{2} ; \boldsymbol{\rho}_{3}, \boldsymbol{\rho}_{4}, 0\right) K^{*}\left(\boldsymbol{r}_{1}, \boldsymbol{\rho}_{1}, z\right)\\ \times K\left(\boldsymbol{r}_{2}, \boldsymbol{\rho}_{2}, z\right) 
 K^{*}\left(\boldsymbol{r}_{3}, \boldsymbol{\rho}_{3}, z\right) K\left(\boldsymbol{r}_{4}, \boldsymbol{\rho}_{4}, z\right) d^{2} \boldsymbol{\rho}_{1} d^{2} \boldsymbol{\rho}_{2} d^{2} \boldsymbol{\rho}_{3} d^{2} \boldsymbol{\rho}_{4},
\end{gathered}
\end{equation}
with the Fresnel propagation kernel defined by
\cite{goodman2005introduction}
\begin{equation}
K(\boldsymbol{r}, \boldsymbol{\rho}, z)=\frac{-i \exp (i k z)}{\lambda z} \exp \left[\frac{i k}{2 z}(\boldsymbol{r}-\boldsymbol{\rho})^{2}\right],
\end{equation}
where $k=2\pi / \lambda$. Interestingly, in the context of the scalar theory, a spatially incoherent source is characterized by means of a delta-correlated intensity function, which indicates that subfields---making up for the whole source---at any two distinct points across the source plane are uncorrelated \cite{mandel1995optical}. In the same spirit, and following previous authors \cite{gori2000use,Saleh05PRL}, I define a partially polarized, spatially incoherent source as one whose four-point correlation matrix elements have the form
\begin{equation}\label{Eq:source}
\begin{aligned}
G_{\alpha \beta \alpha^{\prime} \beta^{\prime}}\left(\boldsymbol{\rho}_{1}, \boldsymbol{\rho}_{2} ; \boldsymbol{\rho}_{3}, \boldsymbol{\rho}_{4}, 0\right)&=\lambda^{4} I_{\alpha \beta}\left(\boldsymbol{\rho}_{1}\right) I_{\alpha^{\prime} \beta^{\prime}}\left(\boldsymbol{\rho}_{3}\right)[\delta\left(\boldsymbol{\rho}_{2}-\boldsymbol{\rho}_{1}\right) \delta\left(\boldsymbol{\rho}_{3}-\boldsymbol{\rho}_{4}\right)\\
&+\delta\left(\boldsymbol{\rho}_{2}-\boldsymbol{\rho}_{3}\right) \delta\left(\boldsymbol{\rho}_{1}-\boldsymbol{\rho}_{4}\right)].
\end{aligned}
\end{equation}
Here, $I_{\alpha\beta}\pare{\boldsymbol{\rho}}$ stands for the intensity, position-dependent, polarized two-photon source function. Note that the sum of delta functions in Eq. (\ref{Eq:source}) is a result of the wavefunction symmetrization due to photon (in)distinguishability \cite{perez2017two}.
By substituting Eq. (\ref{Eq:source}) into Eq. (\ref{Eq:twophoton}) I can thus obtain
\begin{equation}\label{Eq:VCZ}
\begin{aligned}
G_{\alpha \beta \alpha^{\prime} \beta^{\prime}}\left(\boldsymbol{r}_{1}, \boldsymbol{r}_{2} ; \boldsymbol{r}_{3}, \boldsymbol{r}_{4}, z\right)=& \frac{\exp \left[\frac{i k}{2 z}\left(r_{2}^{2}-r_{1}^{2}\right)\right] \exp \left[\frac{i k}{2 z}\left(r_{2}^{4}-r_{1}^{3}\right)\right]}{z^{4}} \iint I_{\alpha \beta}\left(\boldsymbol{\rho}_{1}\right) I_{\alpha^{\prime} \beta^{\prime}}\left(\boldsymbol{\rho}_{3}\right) \\
& \times \exp \left[\frac{-2 \pi i}{\lambda z} \boldsymbol{\rho}_{1} \cdot\left(\boldsymbol{r}_{2}-\boldsymbol{r}_{1}\right)\right] \exp \left[\frac{-2 \pi i}{\lambda z} \boldsymbol{\rho}_{3} \cdot\left(\boldsymbol{r}_{4}-\boldsymbol{r}_{3}\right)\right] d^{2} \boldsymbol{\rho}_{1} d^{2} \boldsymbol{\rho}_{3},
\end{aligned}
\end{equation}
which represents the extension of the van Cittert-Zernike theorem to two-photon, partially polarized fields.
\section{Second-order coherence matrix}

\noindent To show some consequences of the two-photon vectorial van Cittert-Zernike theorem, I now present an example where two-photon correlations are built up during propagation. Starting from a spatially incoherent and unpolarized source, whose four-point correlation matrix is written, in the $\{\ket{HH},\ket{HV},\ket{VH},\ket{VV}\}$ basis, as
\begin{equation} \label{Eq:Gini}
\begin{split}
G_{ini}\pare{x_1,x_2;x_3,x_4,0} & =  j_{ini}\pare{x_1,x_2,0} \otimes j_{ini}\pare{x_3,x_4,0} \\
& = \lambda^{4} I_{0}^{2}\cor{\delta\pare{x_{2}-x_{1}}\delta\pare{x_{3}-x_{4}} + \delta\pare{x_{2}-x_{3}}\delta\pare{x_{1}-x_{4}} } \\
& \hspace{5 mm} \times \begin{bmatrix}
1 & 0 & 0 & 0 \\
0 & 1 & 0 & 0 \\
0 & 0 & 1 & 0 \\
0 & 0 & 0 & 1
\end{bmatrix},
\end{split}
\end{equation}
where
\begin{equation}\label{Eq:jini}
j_{ini}\pare{x_1,x_2,0} = \lambda^{2}I_{0}\delta\pare{x_2 - x_1} \begin{bmatrix}
1 & 0 \\
0 & 1 
\end{bmatrix}
\end{equation}
stands for the two-point correlation matrix for a spatially incoherent and unpolarized photon source \cite{gori2000use}, and $I_{0}$ describes a constant-intensity factor. Note that, for the sake of simplicity, I have restricted the system to a one-dimensional case, i.e., I have taken only one element of the transversal vector $\boldsymbol{r} = (x,y)$.

To polarize the source, I make use of a linear polarizer. Specifically, I cover the source with a linear polarization grating whose angle between its transmission axis and the $x$-axis is a linear function of the form $\theta = \pi x/L$, with $L$ being the length of the grating. The four-point correlation matrix after the polarization grating can thus be written as
\begin{equation}\label{Eq:Gout}
\begin{split}
G_{out}\pare{x_1,x_2;x_3,x_4,0} & =  \cor{P^{\dagger}\pare{x_1}j_{ini}\pare{x_1,x_2,0}P\pare{x_2}} \otimes \cor{P^{\dagger}\pare{x_3}j_{ini}\pare{x_3,x_4,0}P\pare{x_4}} \\
& \times \text{rect}\pare{x_{1}/L}\text{rect}\pare{x_{2}/L}\text{rect}\pare{x_{3}/L}\text{rect}\pare{x_{4}/L} \\
& \times \lambda^{4} I_{0}^{2}\cor{\delta\pare{x_{2}-x_{1}}\delta\pare{x_{3}-x_{4}} + \delta\pare{x_{2}-x_{3}}\delta\pare{x_{1}-x_{4}} }
\end{split}
\end{equation} 
where the product of $\text{rect}\pare{\cdots}$ functions describe the finite size of the source, and the action of the polarization grating is given by the Jones matrix,
\begin{equation}\label{Eq:Jones}
P\pare{x} = \begin{bmatrix}
\cos^{2}\pare{\frac{\pi x}{L}} & \cos\pare{\pi x/p}\sin\pare{\pi x/L} \\
\cos\pare{\pi x/p}\sin\pare{\pi x/L}  & \sin^{2}\pare{\pi x/L} 
\end{bmatrix}.
\end{equation}
By substituting Eqs. (\ref{Eq:Gini})-(\ref{Eq:Jones}) into Eq. (\ref{Eq:VCZ}), I can obtain the explicit form of the polarized, four-point correlation matrix elements. As an example, I can find that, in the far-field---i.e. the region where the quadratic phase factor in front of the integral of Eq. (\ref{Eq:VCZ}) goes to one---the normalized four-point correlation function for H-polarized photons reads as
\begin{equation}
\begin{aligned}
G_{H H H H}&\left(\nu_{1}, \nu_{2}, \nu_{3}, \nu_{4} ; z\right)=\left[\operatorname{sinc}\left(\nu_{1}\right)+\frac{1}{2} \operatorname{sinc}\left(\nu_{1}-1\right)+\frac{1}{2} \operatorname{sinc}\left(\nu_{1}+1\right)\right] \\
& \times\left[\operatorname{sinc}\left(\nu_{2}\right)+\frac{1}{2} \operatorname{sinc}\left(\nu_{2}-1\right)+\frac{1}{2} \operatorname{sinc}\left(\nu_{2}+1\right)\right] \\
&+\frac{1}{16}\left[\operatorname{sinc}\left(2+\nu_{3}\right)\left(\operatorname{sinc}\left(\nu_{4}\right)+2 \operatorname{sinc}\left(1-\nu_{4}\right)+\operatorname{sinc}\left(2-\nu_{4}\right)\right)\right.\\
&+2 \operatorname{sinc}\left(1+\nu_{3}\right)\left(3 \operatorname{sinc}\left(\nu_{4}\right)+3 \operatorname{sinc}\left(1-\nu_{4}\right)+\operatorname{sinc}\left(2-\nu_{4}\right)+\operatorname{sinc}\left(1+\nu_{4}\right)\right) \\
&+\operatorname{sinc}\left(2-\nu_{3}\right)\left(\operatorname{sinc}\left(\nu_{4}\right)+2 \operatorname{sinc}\left(1+\nu_{4}\right)+\operatorname{sinc}\left(2+\nu_{4}\right)\right) \\
&+2 \operatorname{sinc}\left(1-\nu_{3}\right)\left(3 \operatorname{sinc}\left(\nu_{4}\right)+\operatorname{sinc}\left(1-\nu_{4}\right)+3 \operatorname{sinc}\left(1+\nu_{4}\right)+\operatorname{sinc}\left(2+\nu_{4}\right)\right) \\
&\left.+\operatorname{sinc}\left(\nu_{3}\right)\left(10 \operatorname{sinc}\left(\nu_{4}\right)+6 \operatorname{sinc}\left(1-\nu_{4}\right)+\operatorname{sinc}\left(2-\nu_{4}\right)+6 \operatorname{sinc}\left(1+\nu_{4}\right)+\operatorname{sinc}\left(2+\nu_{4}\right)\right)\right]
\end{aligned}
\end{equation}
with
\begin{equation}
\nu_1 = L\frac{x_2-x_3}{\lambda z}; \hspace{2mm} \nu_2 = L\frac{x_4-x_1}{\lambda z}; \hspace{2mm}  \nu_3 = L\frac{x_2-x_1}{\lambda z}; \hspace{2mm} \nu_4 = L\frac{x_4-x_3}{\lambda z}.
\end{equation}
Finally, by realizing that when monitoring the two-photon correlation function with two detectors, at the observation plane in $z$, I must set \cite{Saleh05PRL}: $x_2 = x_3$ and $x_1 = x_4$, I find that
\begin{equation}
G_{HHHH}\pare{\nu_1,\nu_2,\nu_3,\nu_4;z} = G_{HHHH}\pare{0,0,\nu_1,-\nu_1;z}.  
\end{equation}
I can follow the same procedure as above to obtain the remaining terms of the four-point correlation matrix. 
\section{BCP Matrix Elements}

\noindent Each element of the final BCP matrix upon detection is given as follows:
\begin{equation}
\begin{aligned}
g^{(2)}_{\text{HHHH}}(\nu)&=\frac{1}{16} (10 \text{sinc}(\nu)^2+2 (6 \text{sinc}(\nu+1)+\text{sinc}(\nu+2)+6 \text{sinc}(1-\nu)\\&+\text{sinc}(2-\nu)) \text{sinc}(\nu)
+6 \text{sinc}(\nu+1)^2+\text{sinc}(\nu+2)^2+6 \text{sinc}(1-\nu)^2+\text{sinc}(2-\nu)^2\\
&+4 \text{sinc}(\nu+1) \text{sinc}(\nu+2)+4 (\text{sinc}(\nu+1)+\text{sinc}(2-\nu)) \text{sinc}(1-\nu)+16)
\end{aligned}
\end{equation}
\begin{equation}
\begin{aligned}
g^{(2)}_{\text{HHVV}}(\nu)&=g^{(2)}_{\text{VVHH}}(\nu)=\frac{1}{16} (2 \text{sinc}(\nu)^2-2 (\text{sinc}(\nu+2)+\text{sinc}(2-\nu)) \text{sinc}(\nu)\\&+2 (\text{sinc}(1-\nu)
-\text{sinc}(\nu+1))^2+\text{sinc}(\nu+2)^2+\text{sinc}(2-\nu)^2+16)
\end{aligned}
\end{equation}
\begin{equation}
\begin{aligned}
g^{(2)}_{\text{HHVH}}(\nu)&=g^{(2)}_{\text{VHHH}}(\nu)=-\frac{i}{16}  (\text{sinc}(\nu -2)^2+2 \text{sinc}(1 -  \nu ) \text{sinc}(  \nu -2)-\text{sinc}(  \nu +2)^2\\
&-2 \text{sinc}( \nu +2) \text{sinc}(  \nu +1 )+2 (\text{sinc}(1 -  \nu )-\text{sinc}(  \nu +1 )) (\text{sinc}(  \nu )+\text{sinc}(  \nu +1 )\\
&+\text{sinc}(1 - \nu )))\\
&=\left(g^{(2)}_{\text{HHHV}}(\nu)\right)^*=\left( g^{(2)}_{\text{HVHH}}(\nu)\right)^*
\end{aligned}
\end{equation}
\begin{equation}
\begin{aligned}
g^{(2)}_{\text{HVHV}}(\nu)&=g^{(2)}_{\text{VHVH}}(\nu)=\frac{1}{16} (2 \text{sinc}(\nu)^2-2 (\text{sinc}(\nu+2)+\text{sinc}(2-\nu)) \text{sinc}(\nu)\\
&+2 (\text{sinc}(1-\nu)-\text{sinc}(\nu+1))^2+\text{sinc}(\nu+2)^2+\text{sinc}(2-\nu)^2)
\end{aligned}
\end{equation}
\begin{equation}
\begin{aligned}
g^{(2)}_{\text{VHHV}}(\nu)&=g^{(2)}_{\text{HVVH}}(\nu)=\frac{1}{16} (6 \text{sinc}(\nu)^2-2 (\text{sinc}(\nu+2)+\text{sinc}(2-\nu)) \text{sinc}(\nu)\\&
+2 (\text{sinc}(1-\nu)-\text{sinc}(\nu+1))^2-\text{sinc}(\nu+2)^2-\text{sinc}(2-\nu)^2)
\end{aligned}
\end{equation}
\begin{equation}
\begin{aligned}
g^{(2)}_{\text{HVVV}}(\nu)&=g^{(2)}_{\text{VVHV}}(\nu)=\frac{i}{16}  (2 \text{sinc}(\nu+1)^2-2 \text{sinc}(\nu) \text{sinc}(\nu+1)\\
&-2 \text{sinc}(\nu+2) \text{sinc}(\nu+1)+\text{sinc}(\nu+2)^2
-2 \text{sinc}(1-\nu)^2-\text{sinc}(2-\nu)^2\\
&+2 \text{sinc}(\nu) \text{sinc}(1-\nu)+2 \text{sinc}(1-\nu) \text{sinc}(2-\nu)\\
&=\left(g^{(2)}_{\text{VHVV}}(\nu)\right)^*=\left( g^{(2)}_{\text{VVVH}}(\nu)\right)^*
\end{aligned}
\end{equation}
\begin{equation}
\begin{aligned}
g^{(2)}_{\text{VVVV}}(\nu)&= \frac{1}{16} (10 \text{sinc}(\nu)^2+2 (-6 \text{sinc}(\nu+1)+\text{sinc}(\nu+2)-6 \text{sinc}(1-\nu)+\text{sinc}(2-\nu)) \text{sinc}(\nu)\\
&+6 \text{sinc}(\nu+1)^2+\text{sinc}(\nu+2)^2+6 \text{sinc}(1-\nu)^2+\text{sinc}(2-\nu)^2\\
&-4 \text{sinc}(\nu+1) \text{sinc}(\nu+2)+4 (\text{sinc}(\nu+1)-\text{sinc}(2-\nu)) \text{sinc}(1-\nu)+16)
\end{aligned}
\end{equation}
%
\addcontentsline{toc}{chapter}{\bibname}
\begingroup\singlespacing
\bibliography{template.bib}
\bibliographystyle{apa-good}
\endgroup
 %
\clearpage
\singlespacing
\chapter*{\vitaname}
%
\addcontentsline{toc}{chapter}{\vitaname}
\vspace{-1em}
\protect
\protect

\doublespacing
Nathan Miller (born Nathaniel Robert Miller) was born in Rochester, NY.  Thoughout their childhood they expressed a frequent interest in the natural sciences. They would go on to get their Bachelor's at the University of Dayton.  During that time they worked as a research assistant in various fields, but ultimately discovering their love for quantum mechanics under the guidance of Bill Plick.  They would  move on to Louisiana State University and expand their interests into broader quantum technologies.  They hope to apply this knowledge in the future to the development of new quantum technologies. 

\protect
\protect

\end{document}